%&TeX
%%%% NOTE: TYPESET TWICE %%%%%
\input amstex
\documentstyle{amsppt}
%\NoRunningHeads
\leftheadtext{Quantum Boltzmann Equation}\rightheadtext{Benedetto et al.}
\nologo
%%%%%%%%%%%%%%%%%%%%%%%%%%%%%%%%%%%%%%%%%%%%%%%%%%%%%%%%%%%%
%%%%%%%%%%%%%%%% GRECO
%%%%%%%%%%%%%%%%%%%%%%%%%%%%%%%%%%%%%%%%%%%%%%%%%%%%%%%%%%%%
    \let\e=\varepsilon
    
 \let\o=\omega  
  \let\t=\tau

%\let\F=\Phi

%%%%%%%%%%%%%%%%%%%%%%%%%%%%%%%%%%%%%%%%%%%%%%%%%%%%%%%%%%%%
%%%%%%%%%%%%%%%%%%%%% Numerazione pagine
%%%%%%%%%%%%%%%%%%%%%%%%%%%%%%%%%%%%%%%%%%%%%%%%%%%%%%%%%%%%
\def\data{\number\day/\ifcase\month\or gennaio \or febbraio \or marzo \or
aprile \or maggio \or giugno \or luglio \or agosto \or settembre
\or ottobre \or novembre \or dicembre \fi/\number\year;\,\the\time}
\def\dat{\number\day\,\ifcase\month\or {\rm gennaio} \or {\rm febbraio} \or
{\rm marzo}
\or {\rm aprile} \or {\rm maggio} \or {\rm giugno} \or {\rm luglio} \or {\rm
agosto}
\or {\rm settembre} \or {\rm ottobre} \or {\rm novembre} \or {\rm dicembre}
\fi\,\number\year}
\newcount\pgn \pgn=1
\def\foglio{\number\numsec:\number\pgn
\global\advance\pgn by 1}
\def\foglioa{A\number\numsec:\number\pgn
\global\advance\pgn by 1}

\global\newcount\numsec\global\newcount\numfor
%\global\newcount\numfig
\gdef\profonditastruttura{\dp\strutbox}
\def\senondefinito#1{\expandafter\ifx\csname#1\endcsname\relax}
\def\SIA #1,#2,#3 {\senondefinito{#1#2}%
\expandafter\xdef\csname #1#2\endcsname{#3}\else
\write16{\?? ma #1,#2 e' gia' stato definito !!!!} \fi}
\def\etichetta(#1){(\veroparagrafo.\veraformula)%
\SIA e,#1,(\veroparagrafo.\veraformula) %
\global\advance\numfor by 1%
\write15{\string\FU (#1){\equ(#1)}}%
%\write16{ EQ \equ(#1) <==#1 }
}
\def\FU(#1)#2{\SIA fu,#1,#2 }
\def\etichettaa(#1){(A.\veraformula)%
\SIA e,#1,(A.\veraformula) %
\global\advance\numfor by 1%
\write15{\string\FU (#1){\equ(#1)}}%
%\write16{ EQ \equ(#1) <== #1 }
}
\def\getichetta(#1){Fig. \verafigura
\SIA e,#1,{\verafigura} %
\global\advance\numfig by 1%
\write15{\string\FU (#1){\equ(#1)}}%
\write16{ Fig. \equ(#1) ha simbolo #1 }}
\newdimen\gwidth
\def\BOZZA{
\def\alato(##1){%
 {\vtop to \profonditastruttura{\baselineskip
 \profonditastruttura\vss
 \rlap{\kern-\hsize\kern-1.2truecm{$\scriptstyle##1$}}}}}
\def\galato(##1){\gwidth=\hsize \divide\gwidth by 2%
 {\vtop to \profonditastruttura{\baselineskip
 \profonditastruttura\vss
 \rlap{\kern-\gwidth\kern-1.2truecm{$\scriptstyle##1$}}}}}
}
\def\alato(#1){}
\def\galato(#1){}
\def\veroparagrafo{\number\numsec}\def\veraformula{\number\numfor}
\def\verafigura{\number\numfig}
\def\geq(#1){\getichetta(#1)\galato(#1)}
\def\Eq(#1){\eqno{\etichetta(#1)\alato(#1)}}
\def\eq(#1){\etichetta(#1)\alato(#1)}
\def\Eqa(#1){\eqno{\etichettaa(#1)\alato(#1)}}
\def\eqa(#1){\etichettaa(#1)\alato(#1)}
\def\eqv(#1){\senondefinito{fu#1} #1
\write16{#1 non e' (ancora) definito}%
\else\csname fu#1\endcsname\fi}
\def\equ(#1){\senondefinito{e#1}\eqv(#1)\else\csname e#1\endcsname\fi}
\def\include#1{
\openin13=#1.aux \ifeof13 \relax \else
\input #1.aux \closein13 \fi}
\openin14=\jobname.aux \ifeof14 \relax \else
\input \jobname.aux \closein14 \fi
\openout15=\jobname.aux

%%%%%%%%%%%%%%%%%%%%%%%%%%%%%%%%%%%%%%%%%%%%%%%%%%%%%%%%%%%%
%%%%%%%%%%%%%%%%%%%%%%%%%%%%
%%%%%%%%%%%%%%% DEFINIZIONI LOCALI
%%%%%%%%%%%%%%%%%%%%%%%%%%%%%%%%%%%%%%%%%%%%%%%%%%%%%%%%%%%%
 
 \def\\{\noindent}

\let \pt=\partial

\def\tende#1{\vtop{\ialign{##\crcr\rightarrowfill\crcr
 \noalign{\kern-1pt\nointerlineskip}
 \hskip3.pt${\scriptstyle #1}$\hskip3.pt\crcr}}}
\def\otto{{\kern-1.truept\leftarrow\kern-5.truept\to\kern-1.truept}}

\def\mbox{\hbox}

\def\={{\equiv}}

\def\n{\nabla}

\def\var{\varepsilon}

\def\R{\Bbb R}

\def\pa{\partial}

\def\neq {=\hskip-3.2 mm/\hskip1mm}

%%%%%%%%%%%%%%%%%%%%%%%%%%%%%%%%%%%%%%%%%%%%%%%%%%%%%%%%%%%%
%%%%%%%%%%%end of macros%%%%%%%%%
%%%%%%%%%%%%%%%%%%%%%%%%%%%%%%%%%%%%%%%%%%%%%%%%%%%%%%%%%%%%
%\def\reset{\pgn=1\footline={\hss{\tenrm \foglio}\hss}}
\def\reset{}

\def\11{\hbox{l}\!\!\!\hbox{1}\,}

\def\sqr#1#2{{\vcenter{\vbox{\hrule height.#2pt
\hbox{\vrule width.#2pt height #1pt \kern#1pt
\vrule width.#2pt}
\hrule height.#2pt}}}}
\def\square{\mathchoice\sqr56\sqr56\sqr{2.1}3\sqr{1.5}3}
\def\qed{$\square$}

\def\picture #1 by #2 (#3){
 \vbox to #2{
 \hrule width #1 height 0pt depth 0pt
 \vfill
 \special{picture #3}
 }
 }
\def\scaledpicture #1 by #2 (#3 scaled #4){{
 \dimen0=#1 \dimen1=#2
 \divide\dimen0 by 1000 \multiply\dimen0 by #4
 \divide\dimen1 by 1000 \multiply\dimen1 by #4
 \picture \dimen0 by \dimen1 (#3 scaled #4)}
 }

\def\und(#1){$\underline{\hbox{#1}}$}

\def\xp(#1){\hbox{\rm e}^{#1}}

\def\smp{\mathop{{\mathchar"1350}'}}

%\BOZZA

%%%%%%%%%%%%%%%%%%%%%%%%%%%%%%%%%%%%%%%%%%%%%%%%%%%%%%%%%%%%
%MACROS FRANCOIS
%%%%%%%%%%%%%%%%%%%%%%%%%%%%%%%%%%%%%%%%%%%%%%%%%%%%%%%%%%%%

\def\noi{\noindent}
\def\({\Big(}
\def\){\Big)}
%%%%%%%%%%%%%%%%%%%%%%%%%%%%%%%%%%%%%%%%%%%%%%%%%%%%%%%%%%%%

%%%%%%%%%%%%%%%%%%%%%%%%%%%%%%%%%%%%%%%%%%%%%%%%%%%%%%%%%%%%
%%%%%%%%%%%%%%%%%%%%%%%%%%%%%%%%%%%%%%%%%%%%%%%%%%%%%%%%%%%%
%%%%%%%%%%%%%%%%%%%%%%%%%%%%%%%%%%%%%%%%%%%%%%%%%%%%%%%%%%%%
%%%%%%%%%%%%%%%%%%%%%%%%%%%%%%%%%%%%%%%%%%%%%%%%%%%%%%%%%%%%
%%%%%%%%%%%%%%%%%%%%%%%%%%%%%%%%%%%%%%%%%%%%%%%%%%%%%%%%%%%%
%%%%%%%%%%%%%%%%%%%%%%%%%%%%%%%%%%%%%%%%%%%%%%%%%%%%%%%%%%%%

\magnification=1200
\hsize=16.5truecm \vsize=24.truecm
\topmatter
\title
Some Considerations on the derivation \\ of the nonlinear
Quantum
Boltzmann Equation.
\endtitle
\author
D. Benedetto
\footnote"*" {\eightrm Dipartimento di Matematica
Universit\`a di Roma `La Sapienza', P.le A. Moro, 5, 00185 Roma, Italy},
F. Castella
\footnote "$^+$"{\eightrm IRMAR Universit\'e de Rennes 1, Campus de Beaulieu
35042 Rennes-Cedex France},
R. Esposito
\footnote"$^\#$"
{\eightrm Centro Linceo Interdisciplinare ``Beniamino
Segre'', Via della Lungara 10, 00165 Roma,
and Dipar\-timento di Matematica pura ed applicata, Universit\`a di L'Aquila,
Coppito,
67100 L'Aquila, Italy}
and
M. Pulvirenti*
\endauthor

%%%%%%%%%%%%%%%%%%%%%%%%%%%%%%%%%%%%%%%%%%%%%%%%%%%%%%%%%%%%
%%%%%%%%%%%%%%%%%%%%%%%%%%%%%%%%%%%%%%%%%%%%%%%%%%%%%%%%%%%%
%%%%%%%%%%%%%%%%%%%%%%%%%%%%%%%%%%%%%%%%%%%%%%%%%%%%%%%%%%%%
%%%%%%%%%%%%%%%%%%%%%%%%%%%%%%%%%%%%%%%%%%%%%%%%%%%%%%%%%%%%

\abstract
In this paper we analyze a system of $N$ identical quantum particles in a
weak-coupling regime. The time evolution of the Wigner transform of the
one-particle
reduced density matrix is represented by means of a perturbative
series.
The expansion is obtained upon iterating the Duhamel formula.
For short times, we rigorously prove that a subseries of the latter,
converges to the solution of the
Boltzmann
equation which is  physically relevant in the context.
In particular, we recover the transition rate as it is predicted by Fermi's Golden
Rule.
However, we are not able
to prove that the quantity neglected while retaining a subseries of the complete
original
perturbative expansion, indeed vanishes in the limit: we
only give
plausibility arguments in this direction.
The present study holds in any space dimension $d\ge 2$.
\endabstract
\endtopmatter

\heading 1. Introduction.\endheading
\numsec= 1
\reset
\numfor= 1

%%%%%%%%%%%%%%%%%%%%%%%%%%%%%%%%%%%%%%%%%%%%%%%%%%%%%%%%%%%%
%%%%%%%%%%%%%%%%%%%%%%%%%%%%%%%%%%%%%%%%%%%%%%%%%%%%%%%%%%%%
%%%%%%%%%%%%%%%%%%%%%%%%%%%%%%%%%%%%%%%%%%%%%%%%%%%%%%%%%%%%
%%%%%%%%%%%%%%%%%%%%%%%%%%%%%%%%%%%%%%%%%%%%%%%%%%%%%%%%%%%%

A large quantum particle system in a rarefaction regime should be
described by a Boltzmann equation. However, while the rigorous validity
of the Boltzmann equation has been proved for classical systems for short
times [L] or globally in time for special situations [IP] (see Ref. [CIP]
for further comments), there is no rigorous analysis for the equivalent
quantum systems.

The problem is physically relevant because quantum
effects, although usually negligible at ordinary temperatures (except for
few light molecules),
happen to play a r\^ole
in the
applications at mesoscopic level.
We refer, for example,
to the
treatment of electron
gases in semiconductors
(for physical references, see the textbooks [RV], [AM], [Ch], as well
as [Bo], [CTDL] - see also the articles [Fi] or [Co]
- see
[MRS] for a mathematically oriented presentation).
Therefore, establishing a well founded quantum
kinetic
theory is certainly interesting not only from a conceptual
viewpoint but also from a practical one. In fact, kinetic descriptions for
quantum systems, beside dilute gases, include dense weakly interacting
systems, as e.g. the electron gas in semiconductors, whose classical
analogues rather yield diffusion processes.

One pragmatic way to introduce the Quantum Boltzmann equation (see e.g. [CC])
is to solve the
scattering problem in Quantum Mechanics and then to replace, in the classical
Boltzmann
equation, the classical cross section with the Quantum one.

A better logically founded approach is to derive an evolution equation for the
Wigner
transform of a quantum state associated to a dilute particle system. Working
on this equation,
one can hope to recover,
at the quantum level,
the same physical arguments than those used at the classical level
to obtain
propagation of
chaos and a
suitable kinetic description for the one particle distribution
function.
This is the strategy we adopt in the present paper to treat quantum
$N$-particle systems.
We refer to the textbook [CIP], or the article [L] for the analysis
of the
classical case.

\bigskip

Summarizing, in the present paper,
we consider a
quantum $N$-particle system
%(bosons)
and represent the time evolution of the one-particle Wigner function, when
$N\to \infty$ in
the weak-coupling limit, in terms of a perturbative series expansion. On the
basis of some
heuristic arguments developed in the next section, we neglect some terms and
consider only
a subseries which is proven to converge, for short times, to the solution of
the Boltzmann
equation with a suitable cross section computed by quantum rules
(namely the Fermi Golden Rule). Therefore,
the present
analysis is not a rigorous derivation of the Quantum Boltzmann equation, but
we hope it
constitutes a step in this direction.
Our main result is Theorem 3.1 below. Although we work in dimension $3$
the present statements are easily extended
in any space dimension $d \ge 2$.
Our analysis heavily relies on stationary phase computations
(see Proposition 3.3), as well as appropriate representations
of the various solutions of the hierarchies we need to handle.
We leave further comments to the last
section and
conclude the present one by establishing the model and the scaling.

%%%%%%%%%%%%%%%%%%%%%%%%%%%%%%%%%%%%%%%%%%%%%%%%%%%%%%%%%%%%
\bigskip
%%%%%%%%%%%%%%%%%%%%%%%%%%%%%%%%%%%%%%%%%%%%%%%%%%%%%%%%%%%%

We consider a $N$-particle quantum system in $\Bbb R^3$. We
assume the mass of the particles, as well as $\hbar$, to be one. The
interaction is
described by a two-body potential $\phi$ so that the potential energy is:
%%%%%%%%%%%%%%%%%%%%%%%%%%%%%%%%%%%%%%%%%%%%%%%%%%%%%%%%%%%%
$$
U(x_1\dots x_N)=\sum_{i<j} \phi(x_i-x_j).\Eq(0.1)
$$
%%%%%%%%%%%%%%%%%%%%%%%%%%%%%%%%%%%%%%%%%%%%%%%%%%%%%%%%%%%%
The Schr\"odinger equation reads:
%%%%%%%%%%%%%%%%%%%%%%%%%%%%%%%%%%%%%%%%%%%%%%%%%%%%%%%%%%%%
$$
i\pa_t \Psi (X_N,t)=-\frac 12 \Delta_N \Psi (X_N,t)
+U(X_N)\Psi(X_N,t)\Eq(0.2)
$$
%%%%%%%%%%%%%%%%%%%%%%%%%%%%%%%%%%%%%%%%%%%%%%%%%%%%%%%%%%%%
where $\Delta_N=\sum_{i=1}^N \Delta_i$, $\Delta_i$ is the Laplacian with
respect to
the $x_i$ variables, and $X_N$ is a shorthand notation for $x_1\dots x_N$.

We rescale the equation according to the hyperbolic space-time scaling
%%%%%%%%%%%%%%%%%%%%%%%%%%%%%%%%%%%%%%%%%%%%%%%%%%%%%%%%%%%%
$$
x\to \var x
\; ,
\qquad t\to \var t\Eq(0.3)
$$
%%%%%%%%%%%%%%%%%%%%%%%%%%%%%%%%%%%%%%%%%%%%%%%%%%%%%%%%%%%%
and simultaneously we rescale also the potential $\phi\to
 \sqrt\e\phi$.
Hence
the resulting
equation is
%%%%%%%%%%%%%%%%%%%%%%%%%%%%%%%%%%%%%%%%%%%%%%%%%%%%%%%%%%%%
$$
i \var \pa_t \Psi^{\var} (X_N,t)=-\frac {\var^2}{2} \Delta_N \Psi^{\var}
(X_N,t)
+U_\var (X_N)\Psi^{\var}(X_N,t),\Eq(0.4)
$$
%%%%%%%%%%%%%%%%%%%%%%%%%%%%%%%%%%%%%%%%%%%%%%%%%%%%%%%%%%%%
where
%%%%%%%%%%%%%%%%%%%%%%%%%%%%%%%%%%%%%%%%%%%%%%%%%%%%%%%%%%%%
$$
U_{\var}(x_1\dots x_N)=\sum_{i<j} \phi_{\var}(x_i-x_j)\Eq(0.5)
$$
%%%%%%%%%%%%%%%%%%%%%%%%%%%%%%%%%%%%%%%%%%%%%%%%%%%%%%%%%%%%
and
%%%%%%%%%%%%%%%%%%%%%%%%%%%%%%%%%%%%%%%%%%%%%%%%%%%%%%%%%%%%
$$
\phi_\e=\sqrt\e\phi(\frac{x}{\e}).\Eq(0.6)
$$
%%%%%%%%%%%%%%%%%%%%%%%%%%%%%%%%%%%%%%%%%%%%%%%%%%%%%%%%%%%%
Note that $\Psi^\e(X_N,t)$ is fully determined by Eq. \equ(0.4) and the
initial datum which
will be specified later on.

%%%%%%%%%%%%%%%%%%%%%%%%%%%%%%%%%%%%%%%%%%%%%%%%%%%%%%%%%%%%
\bigskip
%%%%%%%%%%%%%%%%%%%%%%%%%%%%%%%%%%%%%%%%%%%%%%%%%%%%%%%%%%%%

We want to analyze the limit
$\e \to 0$ in the above equations,
while keeping
%%%%%%%%%%%%%%%%%%%%%%%%%%%%%%%%%%%%%%%%%%%%%%%%%%%%%%%%%%%%
$$
N=\e^{-3}
\; .
\Eq(weak)
$$
%%%%%%%%%%%%%%%%%%%%%%%%%%%%%%%%%%%%%%%%%%%%%%%%%%%%%%%%%%%%
This kind of limit is usually called weak-coupling limit. Another
possible scaling to be considered is the low-density limit. In this case
$\phi$ is unscaled
but
$N=O(\e^{-2})$. In the classical context this is nothing but the
Boltzmann-Grad limit
(see e.g. [CIP]). In the present paper we will only be concerned with the weak
coupling
limit which is, to some extent,  technically easier.

%%%%%%%%%%%%%%%%%%%%%%%%%%%%%%%%%%%%%%%%%%%%%%%%%%%%%%%%%%%%
\bigskip
%%%%%%%%%%%%%%%%%%%%%%%%%%%%%%%%%%%%%%%%%%%%%%%%%%%%%%%%%%%%

We now introduce the Wigner function:
%%%%%%%%%%%%%%%%%%%%%%%%%%%%%%%%%%%%%%%%%%%%%%%%%%%%%%%%%%%%
$$
W^N(X_N,V_N)=\left(\frac {1}{2\pi}\right)^{3N} \int dY_N \;
e^{iY_N\cdot V_N}
\overline{\Psi}^{\var}(X_N+\frac {\var}{2}Y_N
)\Psi^{\var}(X_N-\frac
{\var}{2} Y_N).\Eq(0.7)
$$
%%%%%%%%%%%%%%%%%%%%%%%%%%%%%%%%%%%%%%%%%%%%%%%%%%%%%%%%%%%%
A standard computation yields:
%%%%%%%%%%%%%%%%%%%%%%%%%%%%%%%%%%%%%%%%%%%%%%%%%%%%%%%%%%%%
$$
(\pa_t+V_N\cdot \nabla_N) W^N(X_N,V_N) =\frac 1{\sqrt\var} \big
(T^\e_NW^N\big)(X_N,V_N)\Eq(0.8)
$$
%%%%%%%%%%%%%%%%%%%%%%%%%%%%%%%%%%%%%%%%%%%%%%%%%%%%%%%%%%%%
where $V_N\cdot\n_N=\sum_{i=1}^N v_i\cdot \n_{x_i}$ and $(\pa_t+V_N\cdot
\nabla_N)$ is the usual
free stream operator.
Also, we have introduced
%%%%%%%%%%%%%%%%%%%%%%%%%%%%%%%%%%%%%%%%%%%%%%%%%%%%%%%%%%%%
$$
(T^\e_NW^N\big)(X_N,V_N)=\sum_{0<k<\ell\le N}
(T^\e_{k,\ell}W^N\big)(X_N,V_N),\Eq(0.9)
$$
%%%%%%%%%%%%%%%%%%%%%%%%%%%%%%%%%%%%%%%%%%%%%%%%%%%%%%%%%%%%
with
%%%%%%%%%%%%%%%%%%%%%%%%%%%%%%%%%%%%%%%%%%%%%%%%%%%%%%%%%%%%
$$
\eqalign{&(T^\e_{k,\ell}W^N\big)(X_N,V_N)=
\frac 1{i}(\frac {1}{2\pi})^{3N} \int
dY_N \int dV_N'
\;
\hbox{\rm e}^{iY_N\cdot (V_N-V_N')}W^N(X_N,V'_N)\cr&
\qquad
\left [\phi\left(\frac {x_k-x_\ell}{\var}-
\frac 12 (y_k-y_\ell) \right)-
\phi \left(\frac {x_k-x_\ell}{\var
}+\frac 12 (y_k-y_\ell)\right)\right]
\; .
}
$$
%%%%%%%%%%%%%%%%%%%%%%%%%%%%%%%%%%%%%%%%%%%%%%%%%%%%%%%%%%%%
In other words,
%%%%%%%%%%%%%%%%%%%%%%%%%%%%%%%%%%%%%%%%%%%%%%%%%%%%%%%%%%%%
$$
\eqalign{&(T^\e_{k,\ell}W^N\big)(X_N,V_N)=
-i\sum_{\sigma=\pm 1}\sigma\int {dh\over (2\pi)^{3}} \hat\phi(h)
\hbox{\rm e}^{i{h\over
\e}(x_k-x_\ell)}
\cr&
\qquad
W^N(x_1,v_1,\dots,x_k,v_k-{\sigma h\over 2},\dots, x_\ell,v_\ell+{\sigma
h\over 2},\dots,x_N,v_N)
.
}\Eq(0.10)
$$
%%%%%%%%%%%%%%%%%%%%%%%%%%%%%%%%%%%%%%%%%%%%%%%%%%%%%%%%%%%%
The operator $T^\e_{k,\ell}$ describes the
``collision''
of particle $k$ with particle $\ell$, and the total operator $T^\e_N$
takes all possible ``collisions'' into account.
Here and below,
$\hat f$ denotes the Fourier transform of $f$,
normalized as
follows:
%%%%%%%%%%%%%%%%%%%%%%%%%%%%%%%%%%%%%%%%%%%%%%%%%%%%%%%%%%%%
$$
\hat f(h)= (\Cal F_x f)(h)= \int_{\Bbb R^3}dx f(x)\hbox{\rm e}^{-ih\cdot
x},\Eq(0.10.2)
$$
%%%%%%%%%%%%%%%%%%%%%%%%%%%%%%%%%%%%%%%%%%%%%%%%%%%%%%%%%%%%
%%%%%%%%%%%%%%%%%%%%%%%%%%%%%%%%%%%%%%%%%%%%%%%%%%%%%%%%%%%%
$$
f(x)= \int_{\Bbb R^3}{dh\over (2\pi)^3}\hat f(h)\hbox{\rm e}^{ih\cdot
x}.\Eq(0.10.5)
$$
%%%%%%%%%%%%%%%%%%%%%%%%%%%%%%%%%%%%%%%%%%%%%%%%%%%%%%%%%%%%

%%%%%%%%%%%%%%%%%%%%%%%%%%%%%%%%%%%%%%%%%%%%%%%%%%%%%%%%%%%%
\bigskip
%%%%%%%%%%%%%%%%%%%%%%%%%%%%%%%%%%%%%%%%%%%%%%%%%%%%%%%%%%%%

We now introduce the Wigner transform of the partial traces according
to the formula, for $j=1.\dots, N-1$:
%%%%%%%%%%%%%%%%%%%%%%%%%%%%%%%%%%%%%%%%%%%%%%%%%%%%%%%%%%%%
$$
f^N_j(X_j,V_j)=\int dx_{j+1} \dots dx_N \int dv_{j+1} \dots dv_N \;
W^N(X_j,x_{j+1} \dots x_N;V_j,v_{j+1} \dots v_N)\Eq(0.11)
$$
%%%%%%%%%%%%%%%%%%%%%%%%%%%%%%%%%%%%%%%%%%%%%%%%%%%%%%%%%%%%
Obviously, we set $f^N_N=W^N$.

\noi From now on we shall suppose that, due to the fact that the
particles are identical, the objects which we have introduced
($\Psi^\var, W^N, f^N_j$) are all symmetric in the exchange of
particles.
%In particular, we assume in this work that we are
%dealing with the dynamics of $N$ bosons.

\noi
Proceeding as in the derivation of the BBKGY hierarchy for classical
systems (see [CIP]),
we
readily arrive
at
the following hierarchy of equations (for $1\leq j\leq
N$):
%%%%%%%%%%%%%%%%%%%%%%%%%%%%%%%%%%%%%%%%%%%%%%%%%%%%%%%%%%%%
$$
(\pa_t+\sum_{k=1}^j v_k\cdot \nabla_k)f^N_j=
\frac {1}{\sqrt {\var}} T^{\var}_j f_j + \frac {N-j}{\sqrt
{\var}}C^{\var}_{j+1}f^N_{j+1}. \Eq(0.12)
$$
%%%%%%%%%%%%%%%%%%%%%%%%%%%%%%%%%%%%%%%%%%%%%%%%%%%%%%%%%%%%
The operator $C^\e_{j+1}$ is defined as:
%%%%%%%%%%%%%%%%%%%%%%%%%%%%%%%%%%%%%%%%%%%%%%%%%%%%%%%%%%%%
$$
C^{\var}_{j+1}=\sum_{k=1}^j C^{\var}_{k,j+1}
\; ,
\Eq(0.13)
$$
%%%%%%%%%%%%%%%%%%%%%%%%%%%%%%%%%%%%%%%%%%%%%%%%%%%%%%%%%%%%
and
%%%%%%%%%%%%%%%%%%%%%%%%%%%%%%%%%%%%%%%%%%%%%%%%%%%%%%%%%%%%
$$
\eqalign{&C^{\var}_{k,j+1}f_{j+1}(x_1\dots x_j;v_1 \dots v_j)=
%\cr&
%\qquad
-i \sum_{\sigma =\pm 1}
\sigma \int {dh \over (2\pi)^3}\int dx_{j+1} \int dv_{j+1} \; \hat \phi (h)
\cr&
\qquad
e^{i \frac {h}{\var} (x_k-x_{j+1})}
\;
f_{j+1}(x_1,x_2,\dots,x_{j+1},v_1,\dots,v_k-\sigma \frac h2, \dots
, v_{j+1}+\sigma \frac h2)
\; .
}
\Eq(0.14)
$$
%%%%%%%%%%%%%%%%%%%%%%%%%%%%%%%%%%%%%%%%%%%%%%%%%%%%%%%%%%%%
%\vfill\eject

\noi The operator $C^\var_{k,j+1}$ describes the
``collision'' of particle $k$, belonging to the $j$-particle
subsystem, with a generic particle outside the subsystem, conventionally
denoted by the number $j+1$ (this numbering uses the fact that all the
particles are identical). The total operator $C^\var_{j+1}$ takes
into account all such collisions. As usual ([CIP]), equation
\equ(0.12) shows that the dynamics of the $j$-particle subsystem
is governed by three effects: the free-stream operator, the
collisions ``inside'' the subsystem (the $T$ term), and the
collisions with particles ``outside'' the subsystem (the $C$
term).

We fix the initial value  $\{f_j^0\}_{j=1}^N$ of the solution
$\{f^N_j(t)\}_{j1}^N$ and we assume for simplicity that
$\{f_j^0\}_{j=1}^N$ is factorized, that is, for all $j=1,N$
%%%%%%%%%%%%%%%%%%%%%%%%%%%%%%%%%%%%%%%%%%%%%%%%%%%%%%%%%%%%
$$
f^0_{j}=f_0^{\otimes j}, \Eq(0.16.1)
$$
%%%%%%%%%%%%%%%%%%%%%%%%%%%%%%%%%%%%%%%%%%%%%%%%%%%%%%%%%%%%
where $f_0$ is a one-particle Wigner function which we assume also
to be a probability distribution. We remind that the quantum
state, whose Wigner transform is a general positive $f_0$, is not
in general a wave function but rather a density matrix. As a consequence
the evolution equation we have to use is not the Schr\"odinger
equation \equ(0.2) but rather the Heisenberg equation for the
density matrix. In both cases the corresponding Wigner equation is
\equ(0.8).

%%%%%%%%%%%%%%%%%%%%%%%%%%%%%%%%%%%%%%%%%%%%%%%%%%%%%%%%%%%%
\bigskip
%%%%%%%%%%%%%%%%%%%%%%%%%%%%%%%%%%%%%%%%%%%%%%%%%%%%%%%%%%%%

One can try to handle the hierarchy \equ(0.12) as for the Boltzmann-Grad limit
for classical
systems, namely to study the asymptotic behavior of the solution expressed in
terms of the
series expansion for $1\le j\le N$, obtained upon iterating the Duhamel
formula,
%%%%%%%%%%%%%%%%%%%%%%%%%%%%%%%%%%%%%%%%%%%%%%%%%%%%%%%%%%%%
$$
\eqalign{&f_j^N(t)=\sum_{n=0}^{N-j}{(N-j)\dots(N-j-n)\over
(\sqrt\e)^n}\int_0^t dt_1\dots\int_0^{t_{n-1}}dt_n  \;
S^\e_{int}(t-t_1)C_{j+1}^\e
\cr&
\qquad
S^\e_{int}(t_1-t_2) C^\e_{j+2}
\dots S^\e_{int}(t_{n-1}-t_n) C^\e_{j+n}
S_{int}^\e(t_n)f^0_{j+n}.}\Eq(0.15)
$$
%%%%%%%%%%%%%%%%%%%%%%%%%%%%%%%%%%%%%%%%%%%%%%%%%%%%%%%%%%%%
Here $S^\e_{int}(t)f_j$ is the $j$-particle interacting flow, namely the
solution to the
initial value problem:
%%%%%%%%%%%%%%%%%%%%%%%%%%%%%%%%%%%%%%%%%%%%%%%%%%%%%%%%%%%%
$$
\cases (\pt_t+V_j\cdot\nabla_j)S^\e_{int}(t)f_j=\displaystyle{1\over
\sqrt\e}T^\e_jS^\e_{int}(t)f_j,\\
S^\e_{int}(0)f_j=f_j.\endcases\Eq(0.16)
$$
%%%%%%%%%%%%%%%%%%%%%%%%%%%%%%%%%%%%%%%%%%%%%%%%%%%%%%%%%%%%

\noi
If we expand $S^\e_{int}(t)$ as a perturbation of the free flow $S(t)$ defined
as
%%%%%%%%%%%%%%%%%%%%%%%%%%%%%%%%%%%%%%%%%%%%%%%%%%%%%%%%%%%%
$$
(S(t) f_j)(X_j,V_j)= f_j(X_j-V_jt, V_j),\Eq(0.17)
$$
%%%%%%%%%%%%%%%%%%%%%%%%%%%%%%%%%%%%%%%%%%%%%%%%%%%%%%%%%%%%
we find
%%%%%%%%%%%%%%%%%%%%%%%%%%%%%%%%%%%%%%%%%%%%%%%%%%%%%%%%%%%%
$$
\eqalign{
&
S^\e_{int}(t)f_j=S(t) f_j+\sum_{m\ge 0}{1\over
(\sqrt\e)^m}\int_0^t d\t_1\int_0^{\t_1}d\t_2 \dots
\int_0^{\t_{m-1}}d\t_m \;
S(t-\t_1)T^\e_j
\cr&
\qquad
\qquad
S(\t_1-\t_2)T^\e_j
\dots
S(\t_{m-1}-\t_m)
T^\e_jS(\t_m)f_j.}
\Eq(0.18)
$$
%%%%%%%%%%%%%%%%%%%%%%%%%%%%%%%%%%%%%%%%%%%%%%%%%%%%%%%%%%%%
Inserting \equ(0.18) into \equ(0.15), the resulting series contains a huge
number of terms. However, we claim that many of these
contributions are
negligible in the limit. In the
next section, we shall give heuristic arguments in that direction.
We first make our point more precise.
We write shortly \equ(0.15) and \equ(0.18) as
%%%%%%%%%%%%%%%%%%%%%%%%%%%%%%%%%%%%%%%%%%%%%%%%%%%%%%%%%%%%
$$
\eqalign{
&
f_j^N
=
\sum_{n=0}^{N-j}
S^\e_{int}
{\widetilde C}^\e_{j+1}
S^\e_{int}
{\widetilde C}^\e_{j+2}
\cdots
S^\e_{int}
{\widetilde C}^\e_{j+n}
S^\e_{int} \; ,
\cr&
S^\e_{int}
=
\sum_{m\ge0}
(S {\widetilde T}_j)^m S
\; ,
}
$$
%%%%%%%%%%%%%%%%%%%%%%%%%%%%%%%%%%%%%%%%%%%%%%%%%%%%%%%%%%%
with an obvious abuse of notation. Here the tilde above $C$ and
$T$ terms absorbs the normalization by $N/\sqrt{\e}$  in
\equ(0.15), respectively $1/\sqrt{\e}$ in \equ(0.18). Now, the
insertion of \equ(0.15) into \equ(0.18) readily gives
%%%%%%%%%%%%%%%%%%%%%%%%%%%%%%%%%%%%%%%%%%%%%%%%%%%%%%%%%%%%
$$
\eqalign
{&
f_j^N
=
\sum_{n=0}^{N-j}
\sum_{m_0 \ge 0}
\sum_{m_1 \ge 0}
\cdots
\sum_{m_n \ge 0}
(S {\widetilde T}_j)^{m_0} S
{\widetilde C}^\e_{j+1}
(S {\widetilde T}_{j+1})^{m_1} S
{\widetilde C}^\e_{j+2}
\cdots
\cr&
\qquad\qquad
\cdots
(S {\widetilde T}_{j+n-1})^{m_{n-1}} S
{\widetilde C}^\e_{j+n}
(S {\widetilde T}_{j+n})^{m_n} S \; .
}
\Eq(expan)
$$
%%%%%%%%%%%%%%%%%%%%%%%%%%%%%%%%%%%%%%%%%%%%%%%%%%%%%%%%%%%%
In view of the expansion \equ(expan) of $f^N_j$, we first claim that
all the
relevant terms
 in \equ(expan) is that corresponding to
%%%%%%%%%%%%%%%%%%%%%%%%%%%%%%%%%%%%%%%%%%%%%%%%%%%%%%%%%%%%
$$
m_0=0 \; , \; m_1=m_2=\cdots=m_n=1 \; .
$$
%%%%%%%%%%%%%%%%%%%%%%%%%%%%%%%%%%%%%%%%%%%%%%%%%%%%%%%%%%%%
Hence as $\e \to 0$, $f_j^N$ is asymptotic to,
%%%%%%%%%%%%%%%%%%%%%%%%%%%%%%%%%%%%%%%%%%%%%%%%%%%%%%%%%%%%
$$
f_j^N \sim
\sum_{n=0}^{N-j}
S {\widetilde C}^\e_{j+1}
S {\widetilde T}_{j+1}
S {\widetilde C}^\e_{j+2}
S {\widetilde T}_{j+2}
\cdots
S {\widetilde T}_{j+n-1}
S {\widetilde C}^\e_{j+n}
S {\widetilde T}_{j+n} S \; .
\Eq(expanbis)
$$
%%%%%%%%%%%%%%%%%%%%%%%%%%%%%%%%%%%%%%%%%%%%%%%%%%%%%%%%%%%%
On the more, expanding each ``collision'' term in \equ(expanbis)
into $C^\e_{j+1}$$=$$\sum_{r=1}^j$$C^\e_{r,j+1}$, and
$T^\e_{j}$$=$ $\sum_{r,\ell=1}^j$$T^\e_{r,\ell}$, we also claim
that $f^N_j$ is in fact asymptotic to,
%%%%%%%%%%%%%%%%%%%%%%%%%%%%%%%%%%%%%%%%%%%%%%%%%%%%%%%%%%%%
$$
f_j^N
\sim
\sum_{n=0}^{N-j}
\sum_{r_1=1}^j
S {\widetilde C}^\e_{r_1,j+1}
S {\widetilde T}_{r_1,j+1}
\sum_{r_2=1}^{j+1}
S {\widetilde C}^\e_{r_2,j+2}
S {\widetilde T}_{r_2,j+2}
\cdots
\sum_{r_n=1}^{j+n-1}
S {\widetilde C}^\e_{r_n,j+n}
S {\widetilde T}_{r_n,j+n} S \; .
\Eq(expanter)
$$
%%%%%%%%%%%%%%%%%%%%%%%%%%%%%%%%%%%%%%%%%%%%%%%%%%%%%%%%%%%%
In other terms, we claim that the dynamics of the $j$-particle subsystem
is only made up of collision/recollision events in the asymptotics $\e\to 0$.

%%%%%%%%%%%%%%%%%%%%%%%%%%%%%%%%%%%%%%%%%%%%%%%%%%%%%%%%%%%%
\bigskip
%%%%%%%%%%%%%%%%%%%%%%%%%%%%%%%%%%%%%%%%%%%%%%%%%%%%%%%%%%%%

Summarizing the above claims, we define the sequence $\{\tilde
f^{N}_j\}_{j=1}^N$ by
%%%%%%%%%%%%%%%%%%%%%%%%%%%%%%%%%%%%%%%%%%%%%%%%%%%%%%%%%%%%
$$
\eqalign{&\tilde f_j^{N}
(t)=S(t) f_j^0+
\sum_{n=1}^{N-j}{(N-j)\dots(N-j-n)\over
(\sqrt\e)^n}\int_0^t dt_1 \int_0^{t_1}d\tau_1
\int_0^{\t_1}dt_2\int_0^{t_2} d\t_2
\dots
\cr&
\qquad
\dots
\int_0^{t_{n-1}}dt_n\int_0^{t_n}d\tau_n
\sum_{r_1=1}^j
\sum_{r_2=1}^{j+1}
\dots
\sum_{r_n=1}^{j+n-1}
S(t-t_1)C_{r_1,j+1}^{\e} S(t_1 - \t_1) T^\e_{r_1,j+1}
\cr&
\qquad\qquad
S(\t_1-t_2)C^{\e}_{r_2,j+2} S(t_2 - \t_2) T^\e_{r_1,j+1}
\dots
\cr&
\qquad\qquad\qquad
\cdots S(\t_{n-1}-t_n)C^{\e}_{r_n,j+n}S(t_n-\t_n) T^\e_{r_n,j+n}
S(\t_n)f^0_{j+n}.}
\Eq(0.19)
$$
%%%%%%%%%%%%%%%%%%%%%%%%%%%%%%%%%%%%%%%%%%%%%%%%%%%%%%%%%%%%
Obviously  $\tilde f_j^N$ is a subseries
of the series expansion
\equ(expan) defining the true value $f^N_j$.
It relates the value of the right-hand-side of \equ(expanter).
The remainder part of this paper is dedicated to the rigorous proof
that
$\tilde f^{\e}_j(t)$ converge to $(f(t))^{\otimes j}$ for
short times, where
$f(t)$ solves the Boltzmann equation with the suitable cross-section.
The
result will be
precisely established and proved in Section 3.

\medskip

\noi
Note that
unfortunately,
we are not able to rigorously prove the asymptotics
\equ(expanter): neither can we prove reasonable uniform bounds on the
relevant series expansions,
nor can we even prove that the asymptotics \equ(expan) holds
{\it term-by-term} .
\bigskip
\bigskip
%%%%%%%%%%%%%%%%%%%%%%%%%%%%%%%%%%%%%%%%%%%%%%%%%%%%%%%%%%%%
%%%%%%%%%%%%%%%%%%%%%%%%%%%%%%%%%%%%%%%%%%%%%%%%%%%%%%%%%%%%
%%%%%%%%%%%%%%%%%%%%%%%%%%%%%%%%%%%%%%%%%%%%%%%%%%%%%%%%%%%%
%%%%%%%%%%%%%%%%%%%%%%%%%%%%%%%%%%%%%%%%%%%%%%%%%%%%%%%%%%%%
\heading 2. Heuristic considerations.\endheading
\numsec= 2
\reset
\numfor= 1

%%%%%%%%%%%%%%%%%%%%%%%%%%%%%%%%%%%%%%%%%%%%%%%%%%%%%%%%%%%%
%%%%%%%%%%%%%%%%%%%%%%%%%%%%%%%%%%%%%%%%%%%%%%%%%%%%%%%%%%%%
%%%%%%%%%%%%%%%%%%%%%%%%%%%%%%%%%%%%%%%%%%%%%%%%%%%%%%%%%%%%
%%%%%%%%%%%%%%%%%%%%%%%%%%%%%%%%%%%%%%%%%%%%%%%%%%%%%%%%%%%%
\bigskip

In this section, we give some reasons justifying the claimed
asymptotics \equ(expanbis) and \equ(expanter). These are
illustrated upon analyzing the lower order terms in the true
expansion \equ(expan). This section thus heuristically justifies
the fact that we restrict ourselves with the mere analysis of
$\tilde f_j^N$ in the present paper.

Also, we analyze the term corresponding to $n=1$ in the expansion
\equ(expanter) defining $\tilde f_j^N$.
We prepare in this way the analysis of the complete series expansion
performed in the next section.

%%%%%%%%%%%%%%%%%%%%%%%%%%%%%%%%%%%%%%%%%%%%%%%%%%%%%%%%%%%%
\bigskip
%%%%%%%%%%%%%%%%%%%%%%%%%%%%%%%%%%%%%%%%%%%%%%%%%%%%%%%%%%%%

To have an idea of the various orders of magnitude of the terms appearing in
\equ(0.15) once
$S^\e_{int}$ is expanded (see \equ(expan)),
we consider the first two terms in \equ(0.15)
expanding
$S^\e_{int}$ up to the first order. As a result we have the following five
terms which we are
going to analyze:
%%%%%%%%%%%%%%%%%%%%%%%%%%%%%%%%%%%%%%%%%%%%%%%%%%%%%%%%%%%%
$$
\Cal I_0=S(t)f^0_j\; ,\Eq(00.1)
$$
%%%%%%%%%%%%%%%%%%%%%%%%%%%%%%%%%%%%%%%%%%%%%%%%%%%%%%%%%%%%
%%%%%%%%%%%%%%%%%%%%%%%%%%%%%%%%%%%%%%%%%%%%%%%%%%%%%%%%%%%%
$$
\Cal I_1={N-j\over
\sqrt\e}\int_0^tdt_1\;S(t-t_1)C^\e_{j+1}S(t_1)f^0_{j+1}\; ,\Eq(00.2)
$$
%%%%%%%%%%%%%%%%%%%%%%%%%%%%%%%%%%%%%%%%%%%%%%%%%%%%%%%%%%%%
%%%%%%%%%%%%%%%%%%%%%%%%%%%%%%%%%%%%%%%%%%%%%%%%%%%%%%%%%%%%
$$
\Cal I_2={1\over
\sqrt\e}\int_0^td\t_1\;S(t-\t_1)T^\e_{j}S(\t_1)f^0_{j}\; ,\Eq(00.3)
$$
%%%%%%%%%%%%%%%%%%%%%%%%%%%%%%%%%%%%%%%%%%%%%%%%%%%%%%%%%%%%
%%%%%%%%%%%%%%%%%%%%%%%%%%%%%%%%%%%%%%%%%%%%%%%%%%%%%%%%%%%%
$$
\Cal
I_3={N-j\over\e}\int_0^td\t_1\int_0^{\t_1}dt_1\;
S(t-\t_1)T^\e_{j}S(\t_1-t_1)C^\e_{j+1}S(t_1)f^0_{j+1}\; ,\Eq(00.4)
$$
%%%%%%%%%%%%%%%%%%%%%%%%%%%%%%%%%%%%%%%%%%%%%%%%%%%%%%%%%%%%
%%%%%%%%%%%%%%%%%%%%%%%%%%%%%%%%%%%%%%%%%%%%%%%%%%%%%%%%%%%%
$$
\eqalign{\Cal I_4&={N-j\over
\sqrt\e}\int_0^tdt_1
\int_0^{t_1}d\t_1 \; S(t-t_1)C^\e_{j+1}S(t_1-\t_1)T^\e_{j+1}S(\t_1)f^0_{j+1}
\cr&
=
\sum_{r=1}^j\sum_{1\le s<\ell\le j+1} \; \Cal I^{r,\ell,s}_4 \; ,}
\Eq(00.5)
$$
%%%%%%%%%%%%%%%%%%%%%%%%%%%%%%%%%%%%%%%%%%%%%%%%%%%%%%%%%%%%
where
%%%%%%%%%%%%%%%%%%%%%%%%%%%%%%%%%%%%%%%%%%%%%%%%%%%%%%%%%%%%
$$
\Cal I^{r,\ell,s}_4={N-j\over
\sqrt\e}\int_0^tdt_1\int_0^{t_1}d\t_1 \;
S(t-t_1)C^{\e}_{r,j+1}S(t_1-\t_1)T^\e_{
\ell,s
}S(\t_1)f^0_{j+1}.\Eq(00.14)
$$
%%%%%%%%%%%%%%%%%%%%%%%%%%%%%%%%%%%%%%%%%%%%%%%%%%%%%%%%%%%%
As we shall see, the terms $\Cal I_i$, $i=1,2,3$ are negligible in the limit $\e\to
0$.  This illustrates (on a particular example) the claim \equ(expanbis)
above.
Also, all the
contributions to $\Cal I_4$ but that for $r=\ell$ and $s=j+1$,
corresponding to a collision/recollision
event,
are equally
vanishing. This illustrates the claim \equ(expanter) above.

%%%%%%%%%%%%%%%%%%%%%%%%%%%%%%%%%%%%%%%%%%%%%%%%%%%%%%%%%%%%
\bigskip
%%%%%%%%%%%%%%%%%%%%%%%%%%%%%%%%%%%%%%%%%%%%%%%%%%%%%%%%%%%%

Clearly $\Cal I_0$ does not require any asymptotic analysis.

%%%%%%%%%%%%%%%%%%%%%%%%%%%%%%%%%%%%%%%%%%%%%%%%%%%%%%%%%%%%
\medskip
%%%%%%%%%%%%%%%%%%%%%%%%%%%%%%%%%%%%%%%%%%%%%%%%%%%%%%%%%%%%

For $\Cal I_1$ we
have:
%%%%%%%%%%%%%%%%%%%%%%%%%%%%%%%%%%%%%%%%%%%%%%%%%%%%%%%%%%%%
$$
\eqalign{&\Cal I_1={N-j\over \sqrt\e} \sum_{r=1}^j\sum_{\sigma=\pm
1}\sigma\int_0^tdt_1\int
dx_{j+1}\int dv_{j+1}
\int {dh\over (2\pi)^3}
\hat\phi(h) \hbox{\rm e}^{i{h\over \e}\cdot\big(x_r-v_r(t-t_1)-x_{j+1}\big)}
\cr&
f_{j+1}^0(x_1-v_1t,\dots,
x_r-v_rt+{\sigma h \over 2}t_1,\dots, x_{j+1}-v_{j+1}t_1-{\sigma h \over
2}t_1,\cr&
\qquad
v_1,\dots,v_r -{\sigma h \over 2},\dots,v_{j+1}) \; .}
\Eq(00.6)
$$
%%%%%%%%%%%%%%%%%%%%%%%%%%%%%%%%%%%%%%%%%%%%%%%%%%%%%%%%%%%%
Now, upon using the stationary phase Theorem for the variables
$x_{j+1}$ and $h$ in formula \equ(00.6), we readily obtain the
following
asymptotics,
valid if $f^0_{j+1}$ and $\phi$ are smooth and decaying enough (which we assume),
%%%%%%%%%%%%%%%%%%%%%%%%%%%%%%%%%%%%%%%%%%%%%%%%%%%%%%%%%%%%
$$
\eqalign{&\Cal I_1={N-j\over \sqrt\e} \times
\Bigg(
i^3 \e^3 \times
\sum_{r=1}^j\sum_{\sigma=\pm
1}\sigma\int_0^tdt_1
\int dv_{j+1}
\hat\phi(0)
\cr&
f_{j+1}^0(x_1-v_1t,\dots,
x_r-v_rt,\dots, x_r-v_r(t-t_1)-v_{j+1}t_1,\cr&
\qquad
v_1,\dots,v_r,\dots,v_{j+1}) + O(\e^4)
\Bigg) \; .
}
\Eq(maini1)
$$
%%%%%%%%%%%%%%%%%%%%%%%%%%%%%%%%%%%%%%%%%%%%%%%%%%%%%%%%%%%%
In other words, the integral in \equ(00.6) concentrates on the set
$x_{j+1}=x_r-v_r(t-t_1)$, $h=0$, as $\e \to 0$, and it has size $\e^3+O(\e^4)$.
Recall indeed that, $\psi(x)$ being a smooth function of $x \in \R^d$,
and $\varphi(x)$ being a  smooth phase such that $\nabla_x
\varphi(x)=0$
iff $x=0$,
and  such that the Hessian at $x=0$, $D^2_{x,x} \varphi(0)$, is
invertible,
then the stationary phase theorem
states the asymptotics (see [H\"o])
%%%%%%%%%%%%%%%%%%%%%%%%%%%%%%%%%%%%%%%%%%%%%%%%%%%%%%%%%%%%
$$
\int_{\R^d}
dx \;
\exp\Big({i \over \e} \varphi(x) \Big)
\psi(x)
=
\e^{{d \over 2}}
{(2i\pi)^{d/2} \over \big( \text{det} D^2_{x,x} \varphi (0)\big)^{1/2}}
\psi(0)
+
O(\e^{1+{d \over 2}}) \; .
$$
%%%%%%%%%%%%%%%%%%%%%%%%%%%%%%%%%%%%%%%%%%%%%%%%%%%%%%%%%%%%
In the case \equ(00.6), we  have $d=6$.
At this stage, we observe that the main contribution in \equ(maini1)
vanishes, due to the sum over $\sigma=\pm 1$. Hence we recover,
%%%%%%%%%%%%%%%%%%%%%%%%%%%%%%%%%%%%%%%%%%%%%%%%%%%%%%%%%%%%
$$
\Cal I_1 = O(N \e^3 {\e \over \e^{1/2}})=O(\e^{1/2}) \; .
$$
%%%%%%%%%%%%%%%%%%%%%%%%%%%%%%%%%%%%%%%%%%%%%%%%%%%%%%%%%%%%

%%%%%%%%%%%%%%%%%%%%%%%%%%%%%%%%%%%%%%%%%%%%%%%%%%%%%%%%%%%%
\medskip
%%%%%%%%%%%%%%%%%%%%%%%%%%%%%%%%%%%%%%%%%%%%%%%%%%%%%%%%%%%%

\noi
For $\Cal I_2$ we have:
%%%%%%%%%%%%%%%%%%%%%%%%%%%%%%%%%%%%%%%%%%%%%%%%%%%%%%%%%%%%
$$
\eqalign{&\Cal I_2={1\over \sqrt\e} \sum_{1\le r<s\le j}\sum_{\sigma=\pm
1}\sigma\int_0^td\t_1 \;
\int {dh\over (2\pi)^3}
\hat\phi( h) \hbox{\rm e}^{i{h\over
\e}\cdot\big((x_r-x_s)-(v_r-v_s)(t-\t_1)\big)}
\cr&
f_{j}^0(x_1-v_1t,\dots,
x_r-v_rt+{\sigma h \over 2}\t_1,\dots,x_s-v_st-{\sigma h \over 2}\t_1,\dots,
x_j-v_jt,\cr&
\qquad
v_1,\dots,v_r -{\sigma h\over 2},\dots,v_s+{\sigma h\over
2},\dots,v_{j}).}
\Eq(00.10)
$$
%%%%%%%%%%%%%%%%%%%%%%%%%%%%%%%%%%%%%%%%%%%%%%%%%%%%%%%%%%%%
Now, we compute the limiting behavior of $\Cal I_2$ in a {\it
weak} sense. Hence we test formula \equ(00.10) against a smooth
function $\psi(x_1,\cdots,x_j,v_1,\cdots,v_j)$. We obtain, using
the stationary phase theorem in the variables $x_r$ and $h$ (one
could use the variables $x_s$ and $h$ as well)
%%%%%%%%%%%%%%%%%%%%%%%%%%%%%%%%%%%%%%%%%%%%%%%%%%%%%%%%%%%%
$$
\int dx_1 \ldots dx_j dv_1 \ldots dv_j \;
\Cal I_2 \; \psi(x_1,\ldots,x_j,v_1,\ldots,v_j)
=
O\({\e^3 \over \e^{1/2}}\)
=O\(\e^{5/2}\) \; .
$$
%%%%%%%%%%%%%%%%%%%%%%%%%%%%%%%%%%%%%%%%%%%%%%%%%%%%%%%%%%%%
This again assumes some smoothness and decay assumptions on $f^0_j$.

%%%%%%%%%%%%%%%%%%%%%%%%%%%%%%%%%%%%%%%%%%%%%%%%%%%%%%%%%%%%
\medskip
%%%%%%%%%%%%%%%%%%%%%%%%%%%%%%%%%%%%%%%%%%%%%%%%%%%%%%%%%%%%

The term $\Cal I_3$ can be treated in the same way.
We write,
%%%%%%%%%%%%%%%%%%%%%%%%%%%%%%%%%%%%%%%%%%%%%%%%%%%%%%%%%%%%
$$
\eqalign{&
\Cal I_3=
{N-j \over \e}
\sum_{1 \leq r < s \leq j}
\sum_{\ell=1}^j
\int_0^t d\t_1 \int_0^{\t_1} dt_1 \;
S(t-\t_1)
T^\e_{r,s}
S(\t_1-t_1)
C^\e_{\ell,j+1}
S(t_1) f^0_{j+1}
\; .
}
\Eq(maini3)
$$
%%%%%%%%%%%%%%%%%%%%%%%%%%%%%%%%%%%%%%%%%%%%%%%%%%%%%%%%%%%%
Here, as in formulae \equ(00.6) and \equ(00.10) defining $\Cal I_1$
and $\Cal I_2$, the $T$ term in  \equ(maini3) induces an integration
over a variable which we call $k_1 \in \R^3$ (playing the r\^ole of
$h$ in formula
\equ(0.10)),
and the $C$ term induces an
integration over variables which we call $h_1 \in \R^3$
(playing the r\^ole of $h$ in formula \equ(0.14)), as well as
$x_{j+1} \in \R^3$ and $v_{j+1} \in \R^3$.
Now, for each value of $r$, $s$, and $\ell$,
the corresponding term in formula
\equ(maini3)
is tested against a smooth function $\psi$, as we did
for $\Cal I_2$. The result involves
an integral over all variables $x_1$, $\dots$,  $x_{j+1}$, $v_1$,
$\ldots$, $v_{j+1}$,
$h_1$ and $k_1$.
As in \equ(00.10),
we now apply the stationary phase theorem. More precisely,
if $r\neq \ell$,
we use the stationary phase in the variables $k_1$ and $x_r$
(to handle the $T$ term), as well as $h_1$ and $x_{j+1}$ (to handle the $C$
term).
In the case $r=\ell$,
we rather use the variables
$k_1$ and $x_s$ for the $T$ term, as well as $h_1$ and $x_{j+1}$ for the $C$ term.
We refer to the treatment of $\Cal I_4$ below for details.
Using this approach, we
eventually obtain,
%%%%%%%%%%%%%%%%%%%%%%%%%%%%%%%%%%%%%%%%%%%%%%%%%%%%%%%%%%%%
$$
\Cal I_3
=
O\Big({N-j \over \e} \e^6\Big)
=
O(\e^2) \; ,
$$
%%%%%%%%%%%%%%%%%%%%%%%%%%%%%%%%%%%%%%%%%%%%%%%%%%%%%%%%%%%%
in a weak sense (i.e. in the sense of distributions in
$x_1$, $\dots$,  $x_{j}$, $v_1$,
$\ldots$, $v_{j}$).
This again requires some smoothness and decay properties for
the function $f^0_{j+1}$ as well as $\phi$.

%%%%%%%%%%%%%%%%%%%%%%%%%%%%%%%%%%%%%%%%%%%%%%%%%%%%%%%%%%%%
\medskip
%%%%%%%%%%%%%%%%%%%%%%%%%%%%%%%%%%%%%%%%%%%%%%%%%%%%%%%%%%%%

To treat the last term $\Cal I_4$ we  have to distinguish various cases:
\item{1)} $\{r, j+1\}\cap\{\ell,s\}=\emptyset$,
\item{2)} $r=\ell$ and $s\neq j+1$,
\item{3)} $r\neq \ell$, $s=j+1$,
\item{4)} $r=\ell$, $s=j+1$,

\noindent and treat them separately.
Item (4) obviously corresponds to
a collision/re\-col\-li\-sion event, and this is the dominant term that we
want to keep in the limit, while replacing the true
expansion \equ(expan) by the reduced series \equ(expanter).

%%%%%%%%%%%%%%%%%%%%%%%%%%%%%%%%%%%%%%%%%%%%%%%%%%%%%%%%%%%%
\medskip
%%%%%%%%%%%%%%%%%%%%%%%%%%%%%%%%%%%%%%%%%%%%%%%%%%%%%%%%%%%%

As regards case 1), we write,
%%%%%%%%%%%%%%%%%%%%%%%%%%%%%%%%%%%%%%%%%%%%%%%%%%%%%%%%%%%%
$$
\eqalign{&\Cal I^{r,\ell,s}_4= {N-j\over
\e}\sum_{\sigma_1,\sigma'_1=\pm 1}
\sigma_1\sigma'_1\int_0^t dt_1\int_0^{t_1}d\t_1\int
dx_{j+1}d
v_{j+1}\int {dh_1\over (2\pi)^3}\int {dk_1\over
(2\pi)^3}
\cr&
\hat\phi(h_1)\hat\phi(k_1)
\hbox{\rm e}^{i{h_1\over
\e}\cdot\big(x_r-x_{j+1}-v_r(t-t_1)\big)}\hbox{\rm e}^{i{k_1\over
\e}\cdot\big(x_\ell-x_s-(v_\ell-v_s)
(t-\t_1)
\big)}
f^0_{j+1}(x_1-v_1t,
\cr&
\qquad
x_\ell-v_\ell
t+{\sigma'_1k_1\over 2}\t_1,\dots,
x_s-v_st-{\sigma'_1k_1\over
2}\t_1,\dots,x_r-v_rt+{\sigma_1h_1\over
2}t_1,\dots,\cr&
\qquad
x_{j+1}-v_{j+1}t-{\sigma_1h_1\over
2}t_1;
v_1,\dots,v_\ell-{\sigma'_1k_1\over 2},\dots,v_s+{\sigma'_1k_1\over 2},\dots,
v_r-{\sigma_1h_1\over 2},\cr&
\qquad\qquad
\dots,v_{j+1}+{\sigma_1h_1\over 2} ).
}
\Eq(00.15)
$$
%%%%%%%%%%%%%%%%%%%%%%%%%%%%%%%%%%%%%%%%%%%%%%%%%%%%%%%%%%%%
As before,
we test formula \equ(00.15) against a smooth function
$\psi$, and
the stationary phase theorem in the variables
$h_1$, $x_{j+1}$ as well as $k_1$, $x_\ell$,
readily gives in this case,
%%%%%%%%%%%%%%%%%%%%%%%%%%%%%%%%%%%%%%%%%%%%%%%%%%%%%%%%%%%%
$$
\Cal I_4^{r,\ell,s}
=
O\(
{N-j \over \e} \e^6
\)
=
O(\e^2) \; ,
$$
%%%%%%%%%%%%%%%%%%%%%%%%%%%%%%%%%%%%%%%%%%%%%%%%%%%%%%%%%%%%
in the sense of distributions in
$x_1$, $\dots$,  $x_{j}$, $v_1$,
$\ldots$, $v_{j}$.

%%%%%%%%%%%%%%%%%%%%%%%%%%%%%%%%%%%%%%%%%%%%%%%%%%%%%%%%%%%%
\medskip
%%%%%%%%%%%%%%%%%%%%%%%%%%%%%%%%%%%%%%%%%%%%%%%%%%%%%%%%%%%%

Case 2) follows along the same lines,
except that we now use the stationary phase Theorem in the variables
$h_1$, $x_{j+1}$, as well as $k_1$, $x_s$,
in the formula,
%%%%%%%%%%%%%%%%%%%%%%%%%%%%%%%%%%%%%%%%%%%%%%%%%%%%%%%%%%%%
$$
\eqalign{&\Cal I^{\ell,\ell,s}_4=
{N-j\over  \e}\sum_{\sigma_1,\sigma'_1=\pm 1}
\sigma_1\sigma'_1\int_0^t dt_1\int_0^{t_1}d\t_1
\int  dx_{j+1}d  v_{j+1}
\int {dh_1\over (2\pi)^3}\int {dk_1\over  (2\pi)^3}
\cr&
\hat\phi(h_1)\hat\phi(k_1)
\hbox{\rm e}^{i{h_1\over \e}\cdot\big(x_\ell-x_{j+1}-v_\ell(t-t_1)\big)}
\hbox{\rm e}^{i{k_1\over \e}\cdot\big(x_\ell-x_s
-(v_\ell-v_s)(t-t_1)
-(v_\ell-{\sigma_1 h_1 \over 2}-v_s)(t_1-\t_1)
\big)}
\cr&
f^0_{j+1}(x_1-v_1t,
x_\ell-v_\ell t
+{\sigma_1 h_1 \over 2}t_1
+{\sigma'_1k_1\over 2}\t_1,\dots,
x_s-v_st-{\sigma'_1k_1\over 2}\t_1,
\dots,\cr&
x_{j+1}-v_{j+1}t-{\sigma_1h_1\over
2}t_1;
v_1,\dots,v_\ell
-
{\sigma_1 h_1\over2}
-
{\sigma'_1k_1\over 2},
\dots,v_s+{\sigma'_1k_1\over 2},
\dots,v_{j+1}+{\sigma_1h_1\over
2}).}
\Eq(i4bis)
$$
%%%%%%%%%%%%%%%%%%%%%%%%%%%%%%%%%%%%%%%%%%%%%%%%%%%%%%%%%%%%

%%%%%%%%%%%%%%%%%%%%%%%%%%%%%%%%%%%%%%%%%%%%%%%%%%%%%%%%%%%%
\medskip
%%%%%%%%%%%%%%%%%%%%%%%%%%%%%%%%%%%%%%%%%%%%%%%%%%%%%%%%%%%%

Case 3) is the same, and we simply need to use stationary phase
in the variables
$h_1$, $x_r$, as well as $k_1$, $x_\ell$,
in a formula analogous to \equ(i4bis).
All these computation require, as usual, some smoothness and decay
assumptions on
$f^0_{j+1}$ and $\phi$.

%%%%%%%%%%%%%%%%%%%%%%%%%%%%%%%%%%%%%%%%%%%%%%%%%%%%%%%%%%%%
\medskip
%%%%%%%%%%%%%%%%%%%%%%%%%%%%%%%%%%%%%%%%%%%%%%%%%%%%%%%%%%%%

Now, case 4) {\it cannot} be treated by a direct stationary phase
approach, since the space variables (the difference $x_\ell-x_{j+1}$)
in the two oscillating factors appearing in this case,
%%%%%%%%%%%%%%%%%%%%%%%%%%%%%%%%%%%%%%%%%%%%%%%%%%%%%%%%%%%%
$$
\eqalign{&
\exp\(i{h_1\over \e}\cdot\big(x_\ell-x_{j+1}-v_\ell(t-t_1)\big)\)
\cr&
\exp\(i{k_1\over  \e}\cdot\big(x_\ell-x_{j+1}
-(v_\ell-v_{j+1})(t-t_1)
-(v_\ell-v_{j+1}-\sigma_1 h_1)(t_1-\t_1)\big)\)
}
\Eq(exosc)
$$
%%%%%%%%%%%%%%%%%%%%%%%%%%%%%%%%%%%%%%%%%%%%%%%%%%%%%%%%%%%%
are the {\it same}: one {\it cannot} decouple the problem into
simply applying twice the stationary phase theorem (one time for each
oscillating factor) as we did before.
It turns out that the present contribution
is, beside $\Cal I_0$,
the only
contribution $O(1)$ which survive in the limit.

%%%%%%%%%%%%%%%%%%%%%%%%%%%%%%%%%%%%%%%%%%%%%%%%%%%%%%%%%%%%
\bigskip
%%%%%%%%%%%%%%%%%%%%%%%%%%%%%%%%%%%%%%%%%%%%%%%%%%%%%%%%%%%%

We thus end up this section by carefully analyzing the limit $\e \to 0$ in
case (4). The analysis performed here allows to understand in a
particular case all the arguments needed for the general proof given in
the next section.
We write,
%%%%%%%%%%%%%%%%%%%%%%%%%%%%%%%%%%%%%%%%%%%%%%%%%%%%%%%%%%%%
$$
\eqalign{&\Cal I^{\ell,\ell,j+1}_4=
{N-j\over  \e}\sum_{\sigma_1,\sigma'_1=\pm 1}
\sigma_1\sigma'_1
\int_0^t dt_1\int_0^{t_1}d\t_1
\int dx_{j+1}d v_{j+1}
\int {dh_1\over (2\pi)^3}\int {dk_1\over (2\pi)^3}
\cr&
\hat\phi(h_1)\hat\phi(k_1)
\hbox{\rm e}^{i{\xi_1\over \e}\cdot\big(x_\ell-x_{j+1}-v_\ell(t-t_1)\big)}
\hbox{\rm e}^{i{k_1\over \e}\cdot\big(x_\ell-x_{j+1}
-(v_\ell-v_{j+1})(t-t_1)-(v_\ell-v_{j+1}-\sigma_1 h_1)(t_1-\t_1)\big)}
\cr&
f^0_{j+1}(x_1-v_1t,
\dots,
x_\ell-v_\ell
t+{\sigma_1 h_1 \over 2}t_1+{\sigma'_1k_1\over 2}\t_1,\dots,
x_{j+1}-v_{j+1}t-{\sigma_1h_1\over
2}t_1-{\sigma'_1 k_1 \over 2} \t_1;\cr&
v_1,\dots,v_\ell-{\sigma_1 h_1 \over 2}-{\sigma'_1k_1\over 2},\dots,
\dots,v_{j+1}+{\sigma_1h_1\over
2}+{\sigma'_1 k_1 \over 2}).}
\Eq(maini4)
$$
%%%%%%%%%%%%%%%%%%%%%%%%%%%%%%%%%%%%%%%%%%%%%%%%%%%%%%%%%%%%
As explained above, one cannot make use of the oscillations of the
product \equ(exosc) by simply using the stationary phase in the
variables
$h_1$, $k_1$, $x_\ell$ and $x_{j+1}$: one has to use the oscillations
in the velocity variable $v_{j+1}$ as well. Unfortunately, the factor
$(t_1-\t_1)$ in \equ(exosc) may vanish, thus killing the oscillation.
This is the very reason for the rescaling we now perform.
We make the following change of variables in \equ(maini4)
%%%%%%%%%%%%%%%%%%%%%%%%%%%%%%%%%%%%%%%%%%%%%%%%%%%%%%%%%%%%
$$
\eqalign{&
t_1-\t_1=\e s_1 \; , \qquad (\text{ i.e. } \; \t_1=t_1-\e s_1 ) \; ,
\cr&
\xi_1=(h_1+k_1)/\e \; .
}
\Eq(chvar)
$$
%%%%%%%%%%%%%%%%%%%%%%%%%%%%%%%%%%%%%%%%%%%%%%%%%%%%%%%%%%%%
The variable $s_1$ is the rescaled time between the first
``collision'' involving particles $\ell$ and $j+1$ (occurring at
time $t_1$), and the recollision event (occurring at time $\t_1$).
This gives in \equ(maini4),
%%%%%%%%%%%%%%%%%%%%%%%%%%%%%%%%%%%%%%%%%%%%%%%%%%%%%%%%%%%%
$$
\eqalign{&\Cal I^{\ell,\ell,j+1}_4=
(N-j)\e^3
\sum_{\sigma_1,\sigma'_1=\pm 1}
\sigma_1\sigma'_1
\int_0^t dt_1\int_0^{t_1/\e}ds_1
\int dx_{j+1}d v_{j+1}
\int {d\xi_1\over (2\pi)^3}\int {dk_1\over (2\pi)^3}
\cr&
\hat\phi(-k_1+\e\xi_1)\hat\phi(k_1)
\hbox{\rm e}^{i\xi_1\cdot\big(x_\ell-x_{j+1}-v_\ell(t-t_1)\big)}
\hbox{\rm e}^{-i s_1 k_1 \cdot
(v_\ell-v_{j+1}-\sigma_1 h_1)}
f^0_{j+1}(\dots) \; .
}
\Eq(i4int)
$$
%%%%%%%%%%%%%%%%%%%%%%%%%%%%%%%%%%%%%%%%%%%%%%%%%%%%%%%%%%%%
Now, in order to treat the limit $\e \to 0$ in \equ(i4int), one
observes that it is of the form,
%%%%%%%%%%%%%%%%%%%%%%%%%%%%%%%%%%%%%%%%%%%%%%%%%%%%%%%%%%%%
$$
A_\e
=
\int_{\R^{4d}}
dx d\xi dy d\eta
\int_{s=0}^{1/\e}
ds \;
\exp(i\xi\cdot x-is \eta\cdot y)
\; \chi(x,y,\xi,\eta) \; ,
\Eq(prot)
$$
%%%%%%%%%%%%%%%%%%%%%%%%%%%%%%%%%%%%%%%%%%%%%%%%%%%%%%%%%%%%
for some smooth function $\chi$.
We claim that,
%%%%%%%%%%%%%%%%%%%%%%%%%%%%%%%%%%%%%%%%%%%%%%%%%%%%%%%%%%%%
$$
A_\e \to (2\pi)^3
\int_{\R^{2d}}
dy d\eta
\int_{s=0}^{+\infty}
ds \;
\exp(-is\eta\cdot y)
\chi(0,0,\xi,\eta) \; ,
\Eq(protlim)
$$
%%%%%%%%%%%%%%%%%%%%%%%%%%%%%%%%%%%%%%%%%%%%%%%%%%%%%%%%%%%%
with the uniform bound,
%%%%%%%%%%%%%%%%%%%%%%%%%%%%%%%%%%%%%%%%%%%%%%%%%%%%%%%%%%%%
$$
\eqalign{&
|A_\e|
\le
C \;
\|\big(\Cal F_{x,y} \chi\big)(\alpha,\beta,\xi,\eta)\|_{L^1(d\xi d\eta;
L^\infty(d\alpha d\beta))}
\cr&
\qquad
\qquad
\qquad
+
C \;
\|\big(\Cal F_{x,y} \chi\big)(\alpha,\beta,\xi,\eta)\|_{L^1(d\alpha d\beta;
L^\infty(d\xi d\eta))}
\; .
}
\Eq(bdosc)
$$
%%%%%%%%%%%%%%%%%%%%%%%%%%%%%%%%%%%%%%%%%%%%%%%%%%%%%%%%%%%%
This is merely a variant of the stationary phase Theorem, and it relies on
the simple two identities,
%%%%%%%%%%%%%%%%%%%%%%%%%%%%%%%%%%%%%%%%%%%%%%%%%%%%%%%%%%%%
$$
\eqalign{&
\int dx d\xi dy d\eta
\;
\hbox{\rm e}^{i x \cdot \xi-is\eta\cdot y}
\chi(x,y,\xi,\eta)
=
\int d\xi d\eta
\(
\Cal F_{x,y} \chi
\)(-\xi,s\eta,\xi,\eta)
\cr&
\qquad
=
s^{-d}
\int d\xi d\eta
\(
\Cal F_{x,y} \chi
\)(-\xi,\eta,\xi,\eta/s)
\; ,
}
\Eq(protman)
$$
%%%%%%%%%%%%%%%%%%%%%%%%%%%%%%%%%%%%%%%%%%%%%%%%%%%%%%%%%%%%
giving, in particular, the absolute convergence of the $ds$ integral
in \equ(protlim).
This model computation
proves that the integral
\equ(i4int)
concentrates asymptotically
on
the set $\xi_1=0$, $x_{j+1}=x_\ell+v_\ell (t-t_1)$:
particles $\ell$ and $j+1$ eventually collide
at the {\it same} point.
The integral over $s_1$ also becomes
an integral over the whole set $\R^+$ as $\e \to 0$,
and all factors $\t_1=t_1-\e s_1$ tend to simply become $\t_1=t_1$:
the two collision/recollision events tend to
take place {\it simultaneously}.
Note the crucial fact that the concentration in the variables
$x_{j+1}$ and $\xi_1$, stemming from the first oscillating factor,
happens independently of the variables $s_1$, $k_1$,
and $v_{j+1}$ of the second oscillating factor.
Hence we recover
the asymptotics,
%%%%%%%%%%%%%%%%%%%%%%%%%%%%%%%%%%%%%%%%%%%%%%%%%%%%%%%%%%%%
$$
\eqalign{& \Cal I^{\ell,\ell,j+1}_4 \sim i^3
\sum_{\sigma_1,\sigma'_1=\pm 1} \sigma_1\sigma'_1 \int_0^t
dt_1\int_0^{+\infty}ds_1 \int d v_{j+1} \int {dk_1\over (2\pi)^3}
\cr& |\hat\phi(k_1)|^2 \exp\(-i s_1 k_1 \cdot
(v_\ell-v_{j+1}+\sigma_1 k_1)\) \cr& f^0_{j+1}(x_1-v_1t, \dots,
x_\ell-v_\ell t-{(\sigma_1 - \sigma'_1)k_1 \over 2}t_1,\dots,
x_{j+1}-v_{j+1}t+{(\sigma_1-\sigma'_1)k_1\over 2}t_1;\cr&
v_1,\dots,v_\ell+{(\sigma_1 - \sigma'_1)k_1 \over 2},\dots,
v_{j+1}-{(\sigma_1-\sigma'_1)k_1\over 2}). } \Eq(maini4fin)
$$
%%%%%%%%%%%%%%%%%%%%%%%%%%%%%%%%%%%%%%%%%%%%%%%%%%%%%%%%%%%%
There remains to use the fact that,
%%%%%%%%%%%%%%%%%%%%%%%%%%%%%%%%%%%%%%%%%%%%%%%%%%%%%%%%%%%%
$$
\hbox{\rm Re} \int_0^\infty ds_1 \; \exp\(-i s_1 k_1 \cdot
(v_\ell-v_{j+1}+\sigma_1 k_1)\) = -\pi \delta\(k_1 \cdot
(v_\ell-v_{j+1}+\sigma_1 k_1) \) \; . \Eq(vpde)
$$
%%%%%%%%%%%%%%%%%%%%%%%%%%%%%%%%%%%%%%%%%%%%%%%%%%%%%%%%%%%%
Using formula \equ(vpde) together with \equ(maini4fin), and
explicitly performing the sum over $\sigma_1$, it turns out that
the term $|\hat\phi(...)|^2$$\exp(...)$ in \equ(maini4fin) becomes
merely
%%%%%%%%%%%%%%%%%%%%%%%%%%%%%%%%%%%%%%%%%%%%%%%%%%%%%%%%%%%%
$$
|\hat\phi(k_1)|^2 \delta\(k_1 \cdot (v_\ell-v_{j+1}+\sigma_1 k_1)
\) \; ,
$$
%%%%%%%%%%%%%%%%%%%%%%%%%%%%%%%%%%%%%%%%%%%%%%%%%%%%%%%%%%%%
as we prove in the next section.
This factor gives the natural cross-section in the limiting
Boltzmann equation: it gives the Fermi Golden Rule together with the
natural conservation of kinetic energy.

%%%%%%%%%%%%%%%%%%%%%%%%%%%%%%%%%%%%%%%%%%%%%%%%%%%%%%%%%%%%
\bigskip
%%%%%%%%%%%%%%%%%%%%%%%%%%%%%%%%%%%%%%%%%%%%%%%%%%%%%%%%%%%%

\noi
As it is clear, the analysis performed in the present paper
heavily relies on the repeated use of the stationary phase Theorem, together
with asymptotic statements of the form
\equ(prot), \equ(protlim), and \equ(bdosc). Eventually, everything boils down to
checking that various oscillating factors involve various {\it independent}
variables,
so as to be able to use that $\exp(i x^2/\e)$, respectively $\exp( i s x^2)$,
has size $\e^{d/2}$,
respectively $s^{-d/2}$, in the sense of distributions in $x \in \R^d$.
This kind of observation is standard in the field, and we may quote
[Sp1],
[EY1], [EY2], [Ca1], [Ca2] for a systematic use.
We also redirect the reader to the next section for precise
statements in the general case.

\bigskip\bigskip
%%%%%%%%%%%%%%%%%%%%%%%%%%%%%%%%%%%%%%%%%%%%%%%%%%%%%%%%%%%%
%%%%%%%%%%%%%%%%%%%%%%%%%%%%%%%%%%%%%%%%%%%%%%%%%%%%%%%%%%%%
%%%%%%%%%%%%%%%%%%%%%%%%%%%%%%%%%%%%%%%%%%%%%%%%%%%%%%%%%%%%
%%%%%%%%%%%%%%%%%%%%%%%%%%%%%%%%%%%%%%%%%%%%%%%%%%%%%%%%%%%%

\heading 3. Convergence. \endheading
\numsec= 3
\reset
\numfor= 1

%%%%%%%%%%%%%%%%%%%%%%%%%%%%%%%%%%%%%%%%%%%%%%%%%%%%%%%%%%%%
%%%%%%%%%%%%%%%%%%%%%%%%%%%%%%%%%%%%%%%%%%%%%%%%%%%%%%%%%%%%
%%%%%%%%%%%%%%%%%%%%%%%%%%%%%%%%%%%%%%%%%%%%%%%%%%%%%%%%%%%%
%%%%%%%%%%%%%%%%%%%%%%%%%%%%%%%%%%%%%%%%%%%%%%%%%%%%%%%%%%%%

\bigskip

The main result of this paper is summarized in the following theorem:
%%%%%%%%%%%%%%%%%%%%%%%%%%%%%%%%%%%%%%%%%%%%%%%%%%%%%%%%%%%%
\proclaim{Theorem 3.1}
Suppose
$\hat\phi \in L^1(\R^3)\cap L^\infty(\R^3)$.
Assume $f_0$ is such that
%%%%%%%%%%%%%%%%%%%%%%%%%%%%%%%%%%%%%%%%%%%%%%%%%%%%%%%%%%%%
$$
\int |\hat f_0(h,k)|dhdk +\int dh\sup_{k\in \Bbb R^3}|\hat
f_0(h,k)|<+\infty.
$$
%%%%%%%%%%%%%%%%%%%%%%%%%%%%%%%%%%%%%%%%%%%%%%%%%%%%%%%%%%%%
Then for all $j\ge 1$ and for $t<t_0$, with $t_0$ sufficiently small depending
on $f_0$ and
$\phi$,
%%%%%%%%%%%%%%%%%%%%%%%%%%%%%%%%%%%%%%%%%%%%%%%%%%%%%%%%%%%%
$$
\lim_{N\to \infty}\tilde f^N_j(t)=f^{\otimes j}(t)
$$
%%%%%%%%%%%%%%%%%%%%%%%%%%%%%%%%%%%%%%%%%%%%%%%%%%%%%%%%%%%%
pointwise in $X_j,V_j,t$.
The function $f(t)$, $t\in [0,t_0)$ is the solution of the
``classical'' Boltzmann equation
%%%%%%%%%%%%%%%%%%%%%%%%%%%%%%%%%%%%%%%%%%%%%%%%%%%%%%%%%%%%
$$
(\pt_t +v\cdot \n_x)f=Q(f,f)
$$
%%%%%%%%%%%%%%%%%%%%%%%%%%%%%%%%%%%%%%%%%%%%%%%%%%%%%%%%%%%%
%%%%%%%%%%%%%%%%%%%%%%%%%%%%%%%%%%%%%%%%%%%%%%%%%%%%%%%%%%%%
$$
Q(f,f)=\int\int dv_1d\o B(\o,|v-v_1|)(f'f'_1-ff_1)
$$
%%%%%%%%%%%%%%%%%%%%%%%%%%%%%%%%%%%%%%%%%%%%%%%%%%%%%%%%%%%%
with $B$ given by
%%%%%%%%%%%%%%%%%%%%%%%%%%%%%%%%%%%%%%%%%%%%%%%%%%%%%%%%%%%%
$$
B(\o,w)=\frac 1{8\pi^2}|\o\cdot w|\,|\hat\phi(\o\,(\o\cdot w))|^2.
$$
%%%%%%%%%%%%%%%%%%%%%%%%%%%%%%%%%%%%%%%%%%%%%%%%%%%%%%%%%%%%
\endproclaim
%%%%%%%%%%%%%%%%%%%%%%%%%%%%%%%%%%%%%%%%%%%%%%%%%%%%%%%%%%%%

%%%%%%%%%%%%%%%%%%%%%%%%%%%%%%%%%%%%%%%%%%%%%%%%%%%%%%%%%%%%
\noi
{\bf Remarks}

\noi
The bounds $\hat f^0$ in $L^1(\R^d;L^1 \cap L^\infty(\R^d))$
in the statement of the Theorem directly stems from estimates of the
type
\equ(bdosc)
above.

\noi
The assumption $\hat\phi \in L^1(\R^3)\cap L^\infty(\R^3)$
is implied by the stronger $\phi \in C^k_0(\R^3)$, whenever $k>3$.

\noi
Finally, the above theorem holds in any space dimension $d \ge 2$, as it is clear
from the proof.
%%%%%%%%%%%%%%%%%%%%%%%%%%%%%%%%%%%%%%%%%%%%%%%%%%%%%%%%%%%%

\medskip

\noindent $\underline{\hbox{Proof}}$: For sake of shortness we
write explicitly the proof for $j=1$. The extension to arbitrary
$j$ is straightforward, as well as the factorization property of
the limit, which follows from Lanford's classical argument (see
[L] and [CIP]).

%%%%%%%%%%%%%%%%%%%%%%%%%%%%%%%%%%%%%%%%%%%%%%%%%%%%%%%%%%%%
\medskip
%%%%%%%%%%%%%%%%%%%%%%%%%%%%%%%%%%%%%%%%%%%%%%%%%%%%%%%%%%%%

\noi
For $j=1$ we can write $f_1^{\e}=\tilde f_1^{N}$ as follows with
$\ell_1=1$
%%%%%%%%%%%%%%%%%%%%%%%%%%%%%%%%%%%%%%%%%%%%%%%%%%%%%%%%%%%%
$$
\eqalign{&
f^{\e}_1 (x_1,v_1,t)
=
\sum_{n= 0}^{N-1} {(N-1)(N-2)\cdots(N-n-1)\over \e^n}
\sum_{\ell_2=1}^2\cdots\sum_{\ell_n=1}^n
\int_0^tdt_1\int_0^{t_1} d\tau_1
\cr&
\int_0^{\t_1} dt_2 \int_0^{t_2} d\t_2
\; \dots
\int_0^{\t_{n-1}}dt_n \int_0^{t_n}  d\tau_n \;
S(t-t_1) C^\e_{\ell_1,2} S(t_1-\t_1) T^{\var}_{\ell_1,2}
\cr&
S(\t_1-t_2) C_{\ell_2,3}^\var S(t_2-\tau_2)T^\e_{\ell_2,2}
\dots
S(\t_{n-1}-t_n)C^\e_{\ell_n,n+1} S(t_n-\tau_n) T^{\var}_{\ell_n,n+1}
S(\tau_n)  f^0_{n+1}.
}
\Eq(1.2)
$$
%%%%%%%%%%%%%%%%%%%%%%%%%%%%%%%%%%%%%%%%%%%%%%%%%%%%%%%%%%%%
The solution to the Boltzmann equation to be compared with the expansion
\equ(1.2) is
%%%%%%%%%%%%%%%%%%%%%%%%%%%%%%%%%%%%%%%%%%%%%%%%%%%%%%%%%%%%
$$
\eqalign{&
f_1 (x_1,v_1,t)=
\sum_{n\ge 0}
\sum_{\ell_2=1}^2\cdots\sum_{\ell_n=1}^n
\int_0^tdt_1 \; \dots \int_0^{t_{n-1}}dt_n
S(t-t_1) C_{\ell_1,2}
\cr&
S(t_1-t_2) C_{\ell_2,3}
\dots
S(t_{n-1}-t_n)C_{\ell_n,n+1}S(t_n)f^0_{n+1},
}
\Eq(1.3)
$$
%%%%%%%%%%%%%%%%%%%%%%%%%%%%%%%%%%%%%%%%%%%%%%%%%%%%%%%%%%%%
with
%%%%%%%%%%%%%%%%%%%%%%%%%%%%%%%%%%%%%%%%%%%%%%%%%%%%%%%%%%%%
$$
\eqalign{&
(C_{\ell,j+1}f_{j+1})
(x_1,v_1,\dots,x_j,v_j)=
\int_{\Bbb R^3}
dv_{j+1}\int_{{\Bbb S}_2} d\o \;
B(\o,w_\ell)
\cr&
\big[f_{j+1}
(x_1,\,\dots,\,x_{\ell},\,\dots,\,x_j,\,x_{\ell};
v_1,\,\dots,\,v_\ell-\o(\o\cdot
w_\ell),\,\dots,\,v_j,\,v_{j+1}+ \o(\o\cdot
w_\ell) )
\cr&
-f_{j+1}
(x_1,\,\dots,\,x_{\ell},\,\dots,\,x_j,\,x_{\ell};
v_1,\,\dots,\,v_\ell,\,\dots,\,v_j,\,v_{j+1}) )
\big],}
\Eq(1.3.5)
$$
%%%%%%%%%%%%%%%%%%%%%%%%%%%%%%%%%%%%%%%%%%%%%%%%%%%%%%%%%%%%
where ${\Bbb S}_2$ is the unit sphere in $\Bbb R^3$,
$w_\ell=v_{\ell}-v_{j+1}$,
and $B(\omega,w)$ is the
quantum cross section computed in the Born approximation.

%%%%%%%%%%%%%%%%%%%%%%%%%%%%%%%%%%%%%%%%%%%%%%%%%%%%%%%%%%%%
\medskip
%%%%%%%%%%%%%%%%%%%%%%%%%%%%%%%%%%%%%%%%%%%%%%%%%%%%%%%%%%%%

Therefore both
$f^\e_1$ and
$f_1$ are expressed as sums of integrals of $f^0_{n+1}$ computed in suitable
points which we
construct in the following way: for a fixed $n$ we call {\it graph of order
$n$ } any sequence
$\ell_1,\dots,\ell_n$ with $\ell_j\in \{1,2,\dots, j\}$ and denote by
$\smp_{\ell_1,\dots,\ell_n}$
the sum on all the possible graphs. Therefore
upon interverting the $t$ and $\tau$ integrals in \equ(1.2) for later
convenience,
%%%%%%%%%%%%%%%%%%%%%%%%%%%%%%%%%%%%%%%%%%%%%%%%%%%%%%%%%%%%
$$
\eqalign{&
f_1^\e(x_1,v_1,t) = \sum_{n\ge
0}\smp_{\ell_1,\dots,\ell_n}{(N-1)(N-2)\cdots(N-n-1)\over \e^n}
(-i)^{2n}
\cr&
\int_0^t dt_1\int_0^{t_1} dt_2
\dots \int_0^{t_{n-1}} dt_n
\;
\int_{t_2}^{t_1} d \tau_1
\int_{t_3}^{t_2} d \tau_2
\dots
\int_{t_n}^{t_{n-1}} d\t_{n-1}
\int_0^{t_n} d\t_n
\cr&
\sum_{\sigma_1,\dots,\sigma_n,\sigma'_1,\dots,\sigma'_n=\pm 1}
\prod_{j=1}^n\sigma_j\sigma'_j
\int {dh_1\over (2\pi)^3}\dots\int {dh_n\over (2\pi)^3}
\int {dk_1\over (2\pi)^3}\dots\int {dk_n\over (2\pi)^3}
\prod_{j=1}^n
\hat \phi(h_j)
\hat \phi(k_j)
\cr&
\int dx_2\dots\int dx_{n+1} \int dv_2\dots\int dv_{n+1}
\; f^0_{n+1}(y_1,\dots,y_{n+1};u_1,\dots,u_{n+1})
\cr&
\prod_{j=1}^n
\exp\big[ i {h_j\over \e}\cdot (y_{\ell_j}(t_j)-y_{j+1}(t_j)) + i {k_j\over
\e}\cdot(y_{\ell_j}(\tau_j)-y_{j+1}(\tau_j)) \big].
}
\Eq(1.4)
$$
%%%%%%%%%%%%%%%%%%%%%%%%%%%%%%%%%%%%%%%%%%%%%%%%%%%%%%%%%%%%
In \equ(1.4) $y_j(s),\ u_j(s)$ are the backward trajectory of the particle $j
\ge 1$ and its velocity.
The particle $j$ is born at time $t_{j-1}$ at $x_j$ with velocity $v_j$
and the trajectory
is
computed
for $s\in[0,t_{j-1}]$,
(as the reader may easily check),
by the following formulae:
we set $t_0=\t_0=t$,
$t_{n+1}=0$ and
%%%%%%%%%%%%%%%%%%%%%%%%%%%%%%%%%%%%%%%%%%%%%%%%%%%%%%%%%%%%
$$
\matrix
u_{j+1}(t_j) = v_{j+1}, \hfill &  \cr
u_{j+1}(\t_j) = u_{j+1}(t_j) + \sigma_j \displaystyle{\frac {h_j}2 }\hfill
& u_{\ell_j}(\t_j) = u_{\ell_j}(t_j) - \sigma_j \displaystyle{\frac
{h_j}2}, \hfill \cr  u_{j+1}(t_{j+1}) = u_{j+1}(\t_j) + \sigma_j'
\displaystyle{\frac {k_j}2} \hfill & u_{l_j}(t_{j+1}) = u_{\ell_j}(\t_j) -
\sigma_j' \displaystyle{\frac {k_j}2}, \hfill \cr  u_r(t_{j+1})
= u_r(\t_j) = u_r(t_j)
\; \text{ if } \; r\neq {\ell_j} \; .
\cr
\endmatrix
\Eq(1.4.1)
$$
%%%%%%%%%%%%%%%%%%%%%%%%%%%%%%%%%%%%%%%%%%%%%%%%%%%%%%%%%%%%
Also, the values of the $u_j$'s are extended outside
the ``collision'' times given by the $t_j$'s and $\t_j$'s,
as follows,
%%%%%%%%%%%%%%%%%%%%%%%%%%%%%%%%%%%%%%%%%%%%%%%%%%%%%%%%%%%%
$$
u_j(s) = \left\{ \matrix
&u_j(\t_r)\phantom{aaa}\hfill &\text{ if }\t_r \le s < t_r, \hfill \cr
&u_j(t_{r+1})\phantom{aaa} \hfill &\text{ if } t_{r+1}\le s < \t_r. \hfill
\endmatrix \right.
\Eq(1.4.2)
$$
%%%%%%%%%%%%%%%%%%%%%%%%%%%%%%%%%%%%%%%%%%%%%%%%%%%%%%%%%%%%
The trajectories $y_j$ themselves are now simply defined by,

%%%%%%%%%%%%%%%%%%%%%%%%%%%%%%%%%%%%%%%%%%%%%%%%%%%%%%%%%%%%
$$
y_j(s) = x_j - \int_s^{t_{j-1}} u_j(\t) \,d\t.
\Eq(1.4.3)
$$
%%%%%%%%%%%%%%%%%%%%%%%%%%%%%%%%%%%%%%%%%%%%%%%%%%%%%%%%%%%%
With this construction $u_j(s)$ is right-continuous, i.e. at the times
$t_r$ and $\t_r$, the velocities are the outgoing ones.
We have also set
$$y_j = y_j(0),\phantom{aaa} u_j=u_j(0).\Eq(1.8)$$

%%%%%%%%%%%%%%%%%%%%%%%%%%%%%%%%%%%%%%%%%%%%%%%%%%%%%%%%%%%%
\medskip
%%%%%%%%%%%%%%%%%%%%%%%%%%%%%%%%%%%%%%%%%%%%%%%%%%%%%%%%%%%%

\noi
We denote by $\Cal T^\e(t_1,\dots,t_n;\ell_1,\dots,\ell_n)$ the contribution
to the $n$-th
order term of the above expansion due to the graph $\{\ell_1,\dots,\ell_n\}$:
%%%%%%%%%%%%%%%%%%%%%%%%%%%%%%%%%%%%%%%%%%%%%%%%%%%%%%%%%%%%
$$
\eqalign{&
\Cal
T^\e(t_1,\dots,t_n;\ell_1,\dots,\ell_n)=
(-i)^{2n}{(N-1)(N-2)\cdots(N-n-1)\over \e^n}
\int_{t_2}^{t_1} d\tau_1 \dots
\cr&
\int_{t_n}^{t_{n-1}} d\tau_{n-1}
\int_{0}^{t_{n}} d\tau_{n}
\sum_{\sigma_1,\dots,\sigma_n,\sigma'_1,\dots,\sigma'_n=\pm 1}
\prod_{j=1}^n\sigma_j\sigma'_j
\cr&
\int {dh_1\over (2\pi)^3}\dots\int {dh_n\over (2\pi)^3}
\int {dk_1\over (2\pi)^3}\dots\int {dk_n\over (2\pi)^3}
\prod_{j=1}^n
\hat \phi(h_j)\hat\phi(k_j)
\cr&
\int dx_2\dots\int dx_{n+1}
\int dv_2\dots\int dv_{n+1} \;
f^0_{n+1}(y_1,\dots,y_{n+1};u_1,\dots,u_{n+1})
\cr&
\prod_{j=1}^n
\exp\big[
i {h_j\over \e}\cdot(y_{\ell_j}(t_j)-y_{j+1}(t_j))
+ i {k_j\over \e}\cdot(y_{\ell_j}(\tau_j)-y_{j+1}(\tau_j))
\big],
}
\Eq(1.9)
$$
%%%%%%%%%%%%%%%%%%%%%%%%%%%%%%%%%%%%%%%%%%%%%%%%%%%%%%%%%%%%
so that
%%%%%%%%%%%%%%%%%%%%%%%%%%%%%%%%%%%%%%%%%%%%%%%%%%%%%%%%%%%%
$$
f_1^\e(t)=\sum_{n\ge 0}\smp_{\ell_1,\dots,\ell_n}\int_0^tdt_1 \dots
\int_0^{t_{n-1}}dt_n\Cal
T^\e(t_1,\dots,t_n;\ell_1,\dots,\ell_n).
\Eq(1.10)
$$
%%%%%%%%%%%%%%%%%%%%%%%%%%%%%%%%%%%%%%%%%%%%%%%%%%%%%%%%%%%%

%%%%%%%%%%%%%%%%%%%%%%%%%%%%%%%%%%%%%%%%%%%%%%%%%%%%%%%%%%%%
\medskip
%%%%%%%%%%%%%%%%%%%%%%%%%%%%%%%%%%%%%%%%%%%%%%%%%%%%%%%%%%%%

\noi
Analogously, we write
%%%%%%%%%%%%%%%%%%%%%%%%%%%%%%%%%%%%%%%%%%%%%%%%%%%%%%%%%%%%
$$
f_1(t)=\sum_{n\ge 0}\smp_{\ell_1,\dots,\ell_n}\int_0^tdt_1 \dots
\int_0^{t_{n-1}}dt_n\Cal T(t_1,\dots,t_n;\ell_1,\dots,\ell_n),
\Eq(1.11)
$$
%%%%%%%%%%%%%%%%%%%%%%%%%%%%%%%%%%%%%%%%%%%%%%%%%%%%%%%%%%%%
with
%%%%%%%%%%%%%%%%%%%%%%%%%%%%%%%%%%%%%%%%%%%%%%%%%%%%%%%%%%%%
$$
\eqalign{&
\Cal T(t_1,\dots,t_n;\ell_1,\dots,\ell_n)
=
\sum_{\sigma \in \{-1,1\}^n}
\int dv_2\dots\int dv_{n+1}
\int_{{\Bbb S}_2} d\omega_2\dots\int_{{\Bbb S}_2} d\omega_{n+1}
\cr&
\big(\prod_{j=1}^n\sigma_j B(\omega_j, w_j)\big)
f^0_{n+1}(y_1^{cl},\dots,y_{n+1}^{cl};u_1^{cl},
\dots,u_{n+1}^{cl}),
}
\Eq(1.12)
$$
%%%%%%%%%%%%%%%%%%%%%%%%%%%%%%%%%%%%%%%%%%%%%%%%%%%%%%%%%%%%
where $w_j=u_{l_j}^{cl}(t_j) - u_{j+1}^{cl}(t_j)$, and the
classical trajectories $y_j^{cl}(s)$ are computed (as the reader
may easily check), by the following formulae: we set $t_0=t$,
$t_{n+1} = 0$, and
%%%%%%%%%%%%%%%%%%%%%%%%%%%%%%%%%%%%%%%%%%%%%%%%%%%%%%%%%%%%
$$
\eqalign{&
u_{j+1}^{cl}(t_j) = v_{j+1},
\cr&
u_{j+1}^{cl}(t_{j+1}) = u_{j+1}^{cl}(t_j)
+ \frac {1+\sigma_j}2 (\omega_j
\cdot (u_{\ell_j}^{cl}(t_j) - u_{j+1}(t_j)) \omega_j
\cr&
u_{\ell_j}^{cl}(t_{j+1}) =
u_{\ell_j}^{cl}(t_j)
- \frac {1+\sigma_j}2 (\omega_j\cdot
(u_{\ell_j}^{cl}(t_j) - u_{j+1}^{cl}(t_j)) \omega_j
\cr&
u_r^{cl}(t_{j+1}) = u_r^{cl}(t_j) \quad \; \text{  if } \; r\neq {\ell_j} \; .
}
\Eq(1.12.1)
$$
%%%%%%%%%%%%%%%%%%%%%%%%%%%%%%%%%%%%%%%%%%%%%%%%%%%%%%%%%%%%
Also, the values of the $u^{cl}_j$'s are extended outside
the ``collision'' times given by the $t_j$'s,
as follows,
%%%%%%%%%%%%%%%%%%%%%%%%%%%%%%%%%%%%%%%%%%%%%%%%%%%%%%%%%%%%
$$
u_j^{cl}(s) = u_j^{cl}(t_{r+1})\phantom{aaa}
\text{ if } t_{r+1}\le s < t_r
\; .
\Eq(1.12.2)
$$
%%%%%%%%%%%%%%%%%%%%%%%%%%%%%%%%%%%%%%%%%%%%%%%%%%%%%%%%%%%%
The trajectories $y^{cl}_j$ themselves are now simply defined by,
%%%%%%%%%%%%%%%%%%%%%%%%%%%%%%%%%%%%%%%%%%%%%%%%%%%%%%%%%%%%
$$
y_j^{cl}(s) = x_j - \int_s^{t_{j-1}} u_j^{cl}(\t) \,d\t.
\Eq(1.12.3)
$$
%%%%%%%%%%%%%%%%%%%%%%%%%%%%%%%%%%%%%%%%%%%%%%%%%%%%%%%%%%%%
Note that, because of the factors $(1+\sigma_r)/2$, when
$\sigma_r=-1$ the velocity is unchanged at time $t_r$, so
producing a loss term, while when $\sigma_r=1$, the velocity
changes by a term such that the conservation of energy is ensured.
Finally,
%%%%%%%%%%%%%%%%%%%%%%%%%%%%%%%%%%%%%%%%%%%%%%%%%%%%%%%%%%%%
$$
y^{cl}_j=y^{cl}_j(0), \quad u^{cl}_j=u^{cl}_j(0).
\Eq(1.18)
$$
%%%%%%%%%%%%%%%%%%%%%%%%%%%%%%%%%%%%%%%%%%%%%%%%%%%%%%%%%%%%

%%%%%%%%%%%%%%%%%%%%%%%%%%%%%%%%%%%%%%%%%%%%%%%%%%%%%%%%%%%%
\medskip
%%%%%%%%%%%%%%%%%%%%%%%%%%%%%%%%%%%%%%%%%%%%%%%%%%%%%%%%%%%%

%%%%%%%%%%%%%%%%%%%%%%%%%%%%%%%%%%%%%%%%%%%%%%%%%%%%%%%%%%%%
\noi
As stated in Theorem 3.1, we want to prove that for all $(x_1,v_1)$,
$$f_1^N(t)\to f_1(t)\Eq(1.19)$$
for $t$ sufficiently small, assuming that:
\item{1)}
$\hat\phi \in L^1(\R^3)\cap L^\infty(\R^3)$ ,
\item{2)} The following norms are finite
%%%%%%%%%%%%%%%%%%%%%%%%%%%%%%%%%%%%%%%%%%%%%%%%%%%%%%%%%%%%
$$
N_1(f^0)=\|\hat f^0\|_{L_1(\Bbb R^3\times\Bbb R^3)},
\quad
N_2(f^0)=\int d\xi\sup_k|\hat f^0(\xi, k)|
\; .
$$
%%%%%%%%%%%%%%%%%%%%%%%%%%%%%%%%%%%%%%%%%%%%%%%%%%%%%%%%%%%%

\noindent This is obtained from two steps: The first step is to
prove that $\Cal T^{\e}$ if uniformly bounded (Proposition 3.2),
the second is to prove Eq. \equ(1.19) via the dominated
convergence theorem,  ensured by Proposition 3.3.
%%%%%%%%%%%%%%%%%%%%%%%%%%%%%%%%%%%%%%%%%%%%%%%%%%%%%%%%%%%%
\medskip
%%%%%%%%%%%%%%%%%%%%%%%%%%%%%%%%%%%%%%%%%%%%%%%%%%%%%%%%%%%%
%%%%%%%%%%%%%%%%%%%%%%%%%%%%%%%%%%%%%%%%%%%%%%%%%%%%%%%%%%%%
\proclaim{Proposition 3.2: (Uniform bound)}
There is $C>0$,
only depending on $\phi$ as in Theorem 3.1,
such that
%%%%%%%%%%%%%%%%%%%%%%%%%%%%%%%%%%%%%%%%%%%%%%%%%%%%%%%%%%%%
$$
|\Cal T^{\e}
(t_1,\dots,t_n;\ell_1,\dots,\ell_n)|\le C^n(N_1(f^0)+N_2(f^0))^n \; .
\Eq(1.20)
$$
%%%%%%%%%%%%%%%%%%%%%%%%%%%%%%%%%%%%%%%%%%%%%%%%%%%%%%%%%%%%
\endproclaim
%%%%%%%%%%%%%%%%%%%%%%%%%%%%%%%%%%%%%%%%%%%%%%%%%%%%%%%%%%%%
\noindent $\underline{\hbox{Proof}}$:
The proof relies essentially on (a systematic use of)
the change of variables \equ(chvar) and the bound \equ(bdosc)
used to treat the term $\Cal I_4^{\ell,\ell,j+1}$ of the previous
section.

First we make
the following change of variables
%%%%%%%%%%%%%%%%%%%%%%%%%%%%%%%%%%%%%%%%%%%%%%%%%%%%%%%%%%%%
$$
\xi_j={h_j+k_j \over \e} \; , \quad s_j={t_j-\tau_j\over \e}.
\Eq(1.21)
$$
%%%%%%%%%%%%%%%%%%%%%%%%%%%%%%%%%%%%%%%%%%%%%%%%%%%%%%%%%%%%
The expression of $T^\e(t_1,\dots,t_n;\ell_1,\dots,\ell_n)$ thus becomes:
%%%%%%%%%%%%%%%%%%%%%%%%%%%%%%%%%%%%%%%%%%%%%%%%%%%%%%%%%%%%
$$
\eqalign{&
\Cal T^\e(t_1,\dots,t_n;\ell_1,\dots,\ell_n)
=
(-i)^{2n}
(N-1)(N-2)\cdots(N-n-1) \; \e^{3n}
\cr&
\int_0^{(t_1-t_2)/\e}ds_1
\int_0^{(t_2-t_3)/\e}ds_2
\; \dots
\int_0^{(t_{n-1}-t_{n})/\e}ds_{n-1}
\int_0^{t_n/\e}ds_n
\sum_{\sigma_1,\dots,\sigma_n,\sigma'_1,\dots,\sigma_n=\pm 1}
\prod_{j=1}^n
\sigma_j\sigma'_j
\cr&
\int {d\xi_1\over (2\pi)^3}\dots\int {d\xi_n\over (2\pi)^3}
\int {dk_1\over (2\pi)^3}\dots\int {dk_n\over (2\pi)^3}
\prod_{j=1}^n\hat\phi(k_j)\hat \phi(-k_j+\e \xi_j)
\cr&
\int dx_2\dots\int dx_{n+1}\int dv_2\dots\int dv_{n+1}  \;
f^0_{n+1}(y_1,\dots,y_{n+1};u_1,\dots,u_{n+1})
\cr&
\prod_{j=1}^n
\exp\big[ i {\xi_j}\cdot (y_{\ell_j}(t_j)-y_{j+1}(t_j)) \big]
\cr&
\prod_{j=1}^n
\exp\big[ - i {k_j\over
\e}\cdot[(y_{\ell_j}(t_j)-y_{j+1}(t_j)) -
(y_{\ell_j}(t_j-\e s_j)-y_{j+1}(t_j-\e s_j))]\big].
}
\Eq(1.22)
$$
%%%%%%%%%%%%%%%%%%%%%%%%%%%%%%%%%%%%%%%%%%%%%%%%%%%%%%%%%%%%
Also,
for fixed $x_1$, $v_1$, and
fixed
$\xi_r, \, h_r, \, s_r,\,t_r,\,\sigma_r,\,\sigma'_r$
($r=1,\dots,n$), we define
the integral
%%%%%%%%%%%%%%%%%%%%%%%%%%%%%%%%%%%%%%%%%%%%%%%%%%%%%%%%%%%%
$$
\eqalign{&
I
=
\int dX dV
\prod_{j=1}^n \hat\phi(k_j)\hat\phi(-k_j+\e\xi_j)
f^0_{n+1}(y_1,\dots,y_{n+1};u_1,\dots,u_{n+1})
\cr&
\prod_{j=1}^n
\exp\big[ i {\xi_j }\cdot(y_{\ell_j}(t_j)-y_{j+1}(t_j)) \big]
\cr&
\prod_{j=1}^n
\exp\big[ - i {k_j\over
\e}\cdot[(y_{\ell_j}(t_j)-y_{j+1}(t_j)) -
(y_{\ell_j}(t_j-\e s_j)-y_{j+1}(t_j-\e s_j))]\big]
\; .
}
\Eq(1.23)
$$
%%%%%%%%%%%%%%%%%%%%%%%%%%%%%%%%%%%%%%%%%%%%%%%%%%%%%%%%%%%%
Here we use the shorthand notation,
%%%%%%%%%%%%%%%%%%%%%%%%%%%%%%%%%%%%%%%%%%%%%%%%%%%%%%%%%%%%
$$
X=(x_2,\dots,x_{n+1}) \; , \;
V=(v_2,\dots,v_{n+1}) \; , \;
\Xi=(\xi_1,\dots,\xi_n) \; , \;
K=(k_1,\dots,k_n) \; .
$$
%%%%%%%%%%%%%%%%%%%%%%%%%%%%%%%%%%%%%%%%%%%%%%%%%%%%%%%%%%%%
In order to bound $\Cal T^\e$ (or $I$), we now follow the same
lines as in the treatment of $\Cal I_4^{\ell,\ell,j+1}$: we use
appropriate changes of variables so as to come up with factors of
the form \equ(prot). Then we treat them as in \equ(protlim),
\equ(bdosc), \equ(protman). The description of the relevant
changes of variables is the reason for the discussion made in
equations \equ(1.24) to \equ(1.35) below. The resulting
Proposition 3.3 below is analogous to the bound \equ(bdosc), and
Proposition 3.4 states a generalization of the asymptotics
\equ(protlim).

%%%%%%%%%%%%%%%%%%%%%%%%%%%%%%%%%%%%%%%%%%%%%%%%%%%%%%%%%%%%
\medskip
%%%%%%%%%%%%%%%%%%%%%%%%%%%%%%%%%%%%%%%%%%%%%%%%%%%%%%%%%%%%

\noi
Let us come to the details.
We note that the mapping
%%%%%%%%%%%%%%%%%%%%%%%%%%%%%%%%%%%%%%%%%%%%%%%%%%%%%%%%%%%%
$$
(x_2,v_2,\dots,x_{n+1},v_{n+1}) \longrightarrow
(y_2,u_2,\dots,y_{n+1},u_{n+1})
\Eq(1.24)
$$
%%%%%%%%%%%%%%%%%%%%%%%%%%%%%%%%%%%%%%%%%%%%%%%%%%%%%%%%%%%%
is one-to-one with unitary Jacobian. Moreover, from Eq.s
\equ(1.4.3),\equ(1.8), we have,
%%%%%%%%%%%%%%%%%%%%%%%%%%%%%%%%%%%%%%%%%%%%%%%%%%%%%%%%%%%%
$$
y_j(s) -( y_j + u_j s) = \int_0^s(u_j(\t) - u_j)\,d\t,
\Eq(1.24.1)
$$
%%%%%%%%%%%%%%%%%%%%%%%%%%%%%%%%%%%%%%%%%%%%%%%%%%%%%%%%%%%%
so that one readily obtains
%%%%%%%%%%%%%%%%%%%%%%%%%%%%%%%%%%%%%%%%%%%%%%%%%%%%%%%%%%%%
$$
y_{\ell_j}(t_j)-y_{j+1}(t_j)
=
y_{\ell_j}-y_{j+1}+(u_{\ell_j}-u_{j+1})t_j
+\gamma^1_j,
\qquad\qquad\qquad\qquad\qquad
\Eq(1.25)
$$
%%%%%%%%%%%%%%%%%%%%%%%%%%%%%%%%%%%%%%%%%%%%%%%%%%%%%%%%%%%%
%%%%%%%%%%%%%%%%%%%%%%%%%%%%%%%%%%%%%%%%%%%%%%%%%%%%%%%%%%%%
$$
y_{\ell_j}(t_j)-y_{j+1}(t_j) -
(y_{\ell_j}(t_j-\e s_j)
-y_{j+1}(t_j-\e s_j))=
\e s_j (u_{\ell_j}-u_{j+1})
+\gamma^2_j
\; .
\Eq(1.26)
$$
%%%%%%%%%%%%%%%%%%%%%%%%%%%%%%%%%%%%%%%%%%%%%%%%%%%%%%%%%%%%
Here $\gamma^i_j$ , $i=1,2$ do not depend on $y_r,\,u_r$ for
$r=2,\,\dots,\,n+1$.
The terms $\gamma_i^j$ actually are linear in the variables $\e \xi_r$ and
$h_r$, for the values $r=j, j+1, \dots, n$.
This gives in \equ(1.23),
%%%%%%%%%%%%%%%%%%%%%%%%%%%%%%%%%%%%%%%%%%%%%%%%%%%%%%%%%%%%
$$
\eqalign{&
I=
\int dY dU \;
\(
\prod_{j=1}^n \hat\phi(k_j)\hat\phi(-k_j+\e \xi_j)
\) \;
\hbox{\rm e}^{i\Gamma( \Xi,K)}
f^0_{1}(y_1,u_1)f^0_{n}(Y,U)
\cr&
\prod_{j=1}^n
\exp\big[ i {\xi_j }\cdot(y_{\ell_j}-y_{j+1}) \big]
\;
\exp\big[
i (u_{\ell_j}-u_{j+1})
\cdot
(-s_j k_j + t_j {\xi_j })
\big]
\; ,
}
\Eq(1.27)
$$
%%%%%%%%%%%%%%%%%%%%%%%%%%%%%%%%%%%%%%%%%%%%%%%%%%%%%%%%%%%%
where
%%%%%%%%%%%%%%%%%%%%%%%%%%%%%%%%%%%%%%%%%%%%%%%%%%%%%%%%%%%%
$$
\Gamma(\Xi,K):=
\sum_{j=1}^n(\gamma_j^1\cdot
\xi_j
-
\e^{-1} \gamma^2_j\cdot k_j)\; ,
\Eq(1.28)
$$
%%%%%%%%%%%%%%%%%%%%%%%%%%%%%%%%%%%%%%%%%%%%%%%%%%%%%%%%%%%%
and we use the shorthand notation
%%%%%%%%%%%%%%%%%%%%%%%%%%%%%%%%%%%%%%%%%%%%%%%%%%%%%%%%%%%%
$$
\eqalign{&
Y= (y_2,\dots,y_{n+1})
\; ,
\quad
U= (u_2,\dots,u_{n+1})
\; .
}
$$
%%%%%%%%%%%%%%%%%%%%%%%%%%%%%%%%%%%%%%%%%%%%%%%%%%%%%%%%%%%%

%%%%%%%%%%%%%%%%%%%%%%%%%%%%%%%%%%%%%%%%%%%%%%%%%%%%%%%%%%%%
\medskip
%%%%%%%%%%%%%%%%%%%%%%%%%%%%%%%%%%%%%%%%%%%%%%%%%%%%%%%%%%%%

\noi Now, as in \equ(protman), we explicitly compute the $dY dU$
integral in \equ(1.27). It is crucial at this stage that
$(y_1,v_1)$ does not depend upon $Y$ nor $U$, and that the phase
$\Gamma/\e$ does not depend on $Y$ nor $U$ neither. Also, we need
the following observation (putting the variables $y_1$ and $u_1$
apart):
%%%%%%%%%%%%%%%%%%%%%%%%%%%%%%%%%%%%%%%%%%%%%%%%%%%%%%%%%%%%
$$
\sum_{j=1}^n \xi_j\cdot (y_{\ell_j}-y_{j+1})= (\Xi,AY)+
\(
\sum_{j=1}^n\delta_{\ell_1,j} \xi_j
\)
\cdot y_1
\; ,
\Eq(1.29)
$$
%%%%%%%%%%%%%%%%%%%%%%%%%%%%%%%%%%%%%%%%%%%%%%%%%%%%%%%%%%%%
%%%%%%%%%%%%%%%%%%%%%%%%%%%%%%%%%%%%%%%%%%%%%%%%%%%%%%%%%%%%
$$
\sum_{j=1}^n \xi_j\cdot (u_{\ell_j}-u_{j+1})t_j= (T \; \Xi,A U)+
\(\sum_{j=1}^n\delta_{\ell_1,j}t_j \xi_j\)\cdot u_1
\; ,
\Eq(1.30)
$$
%%%%%%%%%%%%%%%%%%%%%%%%%%%%%%%%%%%%%%%%%%%%%%%%%%%%%%%%%%%%
%%%%%%%%%%%%%%%%%%%%%%%%%%%%%%%%%%%%%%%%%%%%%%%%%%%%%%%%%%%%
$$
\sum_{j=1}^n s_j k_j\cdot (u_{\ell_j}-u_{j+1}) = ( S \; K,A U)+
\(\sum_{j=1}^n\delta_{\ell_1,j} s_j k_j\)
\cdot u_1
\; ,
\Eq(1.31)
$$
%%%%%%%%%%%%%%%%%%%%%%%%%%%%%%%%%%%%%%%%%%%%%%%%%%%%%%%%%%%%
where $A,\,T ,\, S$ are $n\times n$ matrices, whose elements are
given by:
%%%%%%%%%%%%%%%%%%%%%%%%%%%%%%%%%%%%%%%%%%%%%%%%%%%%%%%%%%%%
$$
A_{r,s}=-\delta_{r,s}+\delta_{{\ell_r},s+1}
\; ,
\Eq(1.32)
$$
%%%%%%%%%%%%%%%%%%%%%%%%%%%%%%%%%%%%%%%%%%%%%%%%%%%%%%%%%%%%
%%%%%%%%%%%%%%%%%%%%%%%%%%%%%%%%%%%%%%%%%%%%%%%%%%%%%%%%%%%%
$$
T=\text{diag}(t_1,\dots,t_n),
\quad S=\text{diag}(s_1,\dots,s_n) \; ,
\; ,
\Eq(1.33)
$$
%%%%%%%%%%%%%%%%%%%%%%%%%%%%%%%%%%%%%%%%%%%%%%%%%%%%%%%%%%%%
and $(\,\cdot\, ,\,\cdot\,)$ denotes the inner product in $\Bbb R^{3n}$.
The matrix $A$ is upper triangular, with $-1$ coefficients on the diagonal.
With these notations we
write,
%%%%%%%%%%%%%%%%%%%%%%%%%%%%%%%%%%%%%%%%%%%%%%%%%%%%%%%%%%%%
$$
\eqalign{&
I=
\int dY dU \;
\(
\prod_{j=1}^n \hat\phi(k_j)\hat\phi(-k_j+\xi_j)
\) \;
\hbox{\rm e}^{i\widetilde\Gamma( \Xi,K)}
f^0_{1}(y_1,u_1)f^0_{n}(Y,U)
\cr&
\qquad
\exp\big[ i {(\Xi, A Y) }  \big]
\;
\exp\big[
- i (S \; K, A U) +i  { (T \; \Xi, A U) }
\big] \; .
}
$$
%%%%%%%%%%%%%%%%%%%%%%%%%%%%%%%%%%%%%%%%%%%%%%%%%%%%%%%%%%%%
Here we have set,
%%%%%%%%%%%%%%%%%%%%%%%%%%%%%%%%%%%%%%%%%%%%%%%%%%%%%%%%%%%%
$$
\widetilde\Gamma( \Xi,K)=
\Gamma( \Xi,K)+ \sum_{j=1}^n\delta_{l_1,j}[\xi_j\cdot
(y_1+t_j u_1)- s_j k_j\cdot
u_1].
\Eq(1.35)
$$
%%%%%%%%%%%%%%%%%%%%%%%%%%%%%%%%%%%%%%%%%%%%%%%%%%%%%%%%%%%%
The $dY dU$ integral is now easily computed,
%%%%%%%%%%%%%%%%%%%%%%%%%%%%%%%%%%%%%%%%%%%%%%%%%%%%%%%%%%%%
$$
\eqalign{&
I=
\hbox{\rm e}^{i\widetilde\Gamma(\e \Xi,K)}
f^0_{1}(y_1,u_1)
\(
\prod_{j=1}^n \hat\phi(k_j)\hat\phi(-k_j+\e \xi_j)
\) \;
{\widehat f}^0_{n}(
-A^{T} \Xi,A^{T} S K - A^{T} T \; \Xi)
\; .
}
\Eq(good)
$$
%%%%%%%%%%%%%%%%%%%%%%%%%%%%%%%%%%%%%%%%%%%%%%%%%%%%%%%%%%%%

%%%%%%%%%%%%%%%%%%%%%%%%%%%%%%%%%%%%%%%%%%%%%%%%%%%%%%%%%%%%
\medskip
%%%%%%%%%%%%%%%%%%%%%%%%%%%%%%%%%%%%%%%%%%%%%%%%%%%%%%%%%%%%

\noi
Armed with expression \equ(good) for $I$,
we are ready to give uniform bounds on this term,
as well as convergence results for $\Cal T^\e$.

%%%%%%%%%%%%%%%%%%%%%%%%%%%%%%%%%%%%%%%%%%%%%%%%%%%%%%%%%%%%
\proclaim{Proposition 3.3}
Let us define the function
%%%%%%%%%%%%%%%%%%%%%%%%%%%%%%%%%%%%%%%%%%%%%%%%%%%%%%%%%%%%
$$
g(\Xi,K)
:=
\Big|
\(
\prod_{j=1}^n \hat\phi(k_j)
\) \;
{\widehat f}^0_{n}(
-A^{T} \Xi,A^{T} S K - A^{T} T \; \Xi)
\Big|
\; ,
$$
%%%%%%%%%%%%%%%%%%%%%%%%%%%%%%%%%%%%%%%%%%%%%%%%%%%%%%%%%%%%
which is independent of $\e$ (the dependence of $g$ upon the $s_j$'s and $t_j$'s
is not made explicit). Then, the following estimate holds true,
%%%%%%%%%%%%%%%%%%%%%%%%%%%%%%%%%%%%%%%%%%%%%%%%%%%%%%%%%%%%
$$
\int_{\R^{6n}}
d\Xi dK \;
g(\Xi,K)
\leq
C^n  \; (N_1(f^0)+N_2(f^0))^n
\; \prod_{j=1}^n
{1 \over (1+ s_j)^3}
\; ,
\Eq(1.37)
$$
%%%%%%%%%%%%%%%%%%%%%%%%%%%%%%%%%%%%%%%%%%%%%%%%%%%%%%%%%%%%
for some constant $C>0$ depending on  $\phi$.
In particular,
$I$ defined by \equ(good) satisfies,
%%%%%%%%%%%%%%%%%%%%%%%%%%%%%%%%%%%%%%%%%%%%%%%%%%%%%%%%%%%%
$$
\Big|
\int_{\R^{6n}}
I \; \;
d\Xi \; dK \;
\Big|
\leq
C^n \|f^0\|_{L^\infty(\R^3)} \; (N_1(f^0)+N_2(f^0))^n
\; \prod_{j=1}^n
{1 \over (1+ s_j)^3}
\; .
$$
%%%%%%%%%%%%%%%%%%%%%%%%%%%%%%%%%%%%%%%%%%%%%%%%%%%%%%%%%%%%
 Hence $I$
is uniformly (in $\e$) integrable
with respect to the variables $s_1 \ge 0$, $\dots$, $s_n \ge 0$.
\endproclaim
%%%%%%%%%%%%%%%%%%%%%%%%%%%%%%%%%%%%%%%%%%%%%%%%%%%%%%%%%%%%
%%%%%%%%%%%%%%%%%%%%%%%%%%%%%%%%%%%%%%%%%%%%%%%%%%%%%%%%%%%%
\noi
{\bf Remark}

\noi
If the space dimension is $d$, the factors
$(1+s_j)^{-3}$ above become
$(1+s_j)^{-d}$.

\medskip

\noindent
$\underline{\hbox{Proof}}$:

The following first bound obviously follows from
\equ(good), together with the fact that
$|\det A|=1$:
%%%%%%%%%%%%%%%%%%%%%%%%%%%%%%%%%%%%%%%%%%%%%%%%%%%%%%%%%%%%
$$
\eqalign{&
\int d\Xi dK \; g(\Xi,K)
\le
C \;
\|\hat \phi\|^n_{L^1(\R^3)}
\;
\|\widehat f^0_n\|_{L^1(\R^d;L^\infty(\R^d))} \; .
}
$$
%%%%%%%%%%%%%%%%%%%%%%%%%%%%%%%%%%%%%%%%%%%%%%%%%%%%%%%%%%%%
On the other hand, to recover decay estimates as the $s$-variables grow, one
makes the change of variables $K \to S^{-1} K$ in \equ(good)
(valid if the $s_i$'s are, say $\neq 0$), and we get,
%%%%%%%%%%%%%%%%%%%%%%%%%%%%%%%%%%%%%%%%%%%%%%%%%%%%%%%%%%%%
$$
\eqalign{&
\int d\Xi dK \; g(\Xi,K)
\le
\(\prod_{j=1}^n s_j^{-3}\)
\int d\Xi \;  dK \;
\(
\prod_{j=1}^n \big|\hat\phi\({k_j \over s_j}\)\big|
\)
|{\widehat f}^0_{n}(
-A^{T} \Xi,A^{T} K - A^{T} T \; \Xi)|
\cr&
\le
C \;
\(\prod_{j=1}^n s_j^{-3}\)
\;
\|\hat \phi\|^{n}_{L^\infty(\R^3)}
\;
\|\widehat f^0_n\|_{L^1(\R^d;L^1(\R^d))} \; .
}
$$
%%%%%%%%%%%%%%%%%%%%%%%%%%%%%%%%%%%%%%%%%%%%%%%%%%%%%%%%%%%%
All this proves the proposition.

\hfill\qed
%%%%%%%%%%%%%%%%%%%%%%%%%%%%%%%%%%%%%%%%%%%%%%%%%%%%%%%%%%%%

%%%%%%%%%%%%%%%%%%%%%%%%%%%%%%%%%%%%%%%%%%%%%%%%%%%%%%%%%%%%
\bigskip
%%%%%%%%%%%%%%%%%%%%%%%%%%%%%%%%%%%%%%%%%%%%%%%%%%%%%%%%%%%%

Using the Proposition 3.3 we conclude the proof of the uniform bound
\equ(1.20): the
factor
%%%%%%%%%%%%%%%%%%%%%%%%%%%%%%%%%%%%%%%%%%%%%%%%%%%%%%%%%%%%
$$
(N-1)(N-2)\cdots(N-n-1) \e^{3n}
$$
%%%%%%%%%%%%%%%%%%%%%%%%%%%%%%%%%%%%%%%%%%%%%%%%%%%%%%%%%%%%
in \equ(1.22)
is bounded by $c^n$,
and the number of terms in $\sum_{\sigma,\sigma'}$ is
$2^{2n}$.
The bound \equ(1.20) implies the absolute uniform convergence of the series
\equ(1.4) for
sufficiently small times. Indeed, the number of graphs
$\{\ell_1,\dots,\ell_n\}$ is $n!$,
while the time ordered integral gives a factor $t^n/n!$. Therefore we have
%%%%%%%%%%%%%%%%%%%%%%%%%%%%%%%%%%%%%%%%%%%%%%%%%%%%%%%%%%%%
$$
|f_1(t;x_1, v_1)|\le \sum_{n=0}^\infty [C(N_1(f^0)+N_2(f^0))t]^n
\; ,
\Eq(1.41)
$$
%%%%%%%%%%%%%%%%%%%%%%%%%%%%%%%%%%%%%%%%%%%%%%%%%%%%%%%%%%%%
converging for $C(N_1(f^0)+N_2(f^0))t<1$.

\hfill
\qed
%%%%%%%%%%%%%%%%%%%%%%%%%%%%%%%%%%%%%%%%%%%%%%%%%%%%%%%%%%%%

%%%%%%%%%%%%%%%%%%%%%%%%%%%%%%%%%%%%%%%%%%%%%%%%%%%%%%%%%%%%
\bigskip
%%%%%%%%%%%%%%%%%%%%%%%%%%%%%%%%%%%%%%%%%%%%%%%%%%%%%%%%%%%%

\noindent
%%%%%%%%%%%%%%%%%%%%%%%%%%%%%%%%%%%%%%%%%%%%%%%%%%%%%%%%%%%%
\proclaim{Proposition 3.4 (Term by term convergence)}
For all $(x_1,v_1)$, and any choice of $t>t_1>\dots>t_n>0$ and
$\{\ell_1,\dots,\ell_n\}$
with $\ell_j\in \{1,2,\dots,j\}$,
%%%%%%%%%%%%%%%%%%%%%%%%%%%%%%%%%%%%%%%%%%%%%%%%%%%%%%%%%%%%
$$
\lim_{\e\to 0} \Cal
T^\e(t_1,\dots,t_n;\ell_1,\dots,\ell_n)=\Cal
T(t_1,\dots,t_n;\ell_1,\dots,\ell_n)
\; .
\Eq(1.42)
$$
%%%%%%%%%%%%%%%%%%%%%%%%%%%%%%%%%%%%%%%%%%%%%%%%%%%%%%%%%%%%
\endproclaim
%%%%%%%%%%%%%%%%%%%%%%%%%%%%%%%%%%%%%%%%%%%%%%%%%%%%%%%%%%%%
\noindent$\underline{\hbox{Proof}}$:
We want to pass to the limit $\e \to 0$ in formula \equ(1.22)
giving the value of $\Cal T^\e$.
Now, equation \equ(1.22) expresses $\Cal T^\e$ as the integral
over $s_1$, $\dots$, $s_n$, together with $\Xi$ and $K$, of the quantity $I$ used
above.
Hence
Proposition 3.3 makes it possible to
use the {\it
dominated convergence} theorem, and we can safely
interchange the $\lim_{\e\to 0}$ with
the integration
on
$(s_1,\dots,s_n)$, and $\Xi$, $K$ in \equ (1.22).

\medskip

\noi
Coming back to formula \equ(1.23) expressing $I$ as an integral over variables $X$
and $V$,
we may use that
the function $f^0_{n+1}(Y,U)$
(seen as a function of $X$ and $V$) is
a fixed (independent of $\e$) function in $L^1(\R^{6n})$.
Hence we can again use the dominated convergence Theorem to
interchange $\lim_{\e\to 0}$ with the
integration on $X$ and $V$ in \equ(1.23).

\medskip

\noi From Eq.s \equ(1.4.1)-\equ(1.4.3) and using $h_j = -k_j + \e
\xi_j$, we have that $y_j(s) \to \bar y_j(s)$, and, for $s\neq
t_r$, $u_j(s) \to \bar u_j(s)$, where $\bar u_j(s)$ are defined in
a similar way of $u_j^{cl}(s)$, substituting the recursive
relation \equ(1.12.1) with
%%%%%%%%%%%%%%%%%%%%%%%%%%%%%%%%%%%%%%%%%%%%%%%%%%%%%%%%%%%%
$$
\eqalign{&
\bar u_{j+1}(t_{j+1}) = \bar u_{j+1}(t_j) + \frac {\sigma'_j - \sigma_j}2
k_j \cr
&\bar u_{\ell_j}(t_{j+1}) =\bar u_{\ell_j}(t_j) -
\frac {\sigma'_j - \sigma_j}2 k_j
\; .
}
\Eq(1.42.10)
$$
%%%%%%%%%%%%%%%%%%%%%%%%%%%%%%%%%%%%%%%%%%%%%%%%%%%%%%%%%%%%
The trajectories are given by
%%%%%%%%%%%%%%%%%%%%%%%%%%%%%%%%%%%%%%%%%%%%%%%%%%%%%%%%%%%%
$$
\bar y_j(s) = x_j - \int_s^{t_{j-1}} \bar u_j(\t) \,d\t \; .
\Eq(1.42.11)
$$
%%%%%%%%%%%%%%%%%%%%%%%%%%%%%%%%%%%%%%%%%%%%%%%%%%%%%%%%%%%%
Moreover, from \equ(1.4.1) and the position $h_j = -k_j + \e \xi_j$
%%%%%%%%%%%%%%%%%%%%%%%%%%%%%%%%%%%%%%%%%%%%%%%%%%%%%%%%%%%%
$$
\eqalign{&
\frac {y_{\ell_j}(t_j)- y_{\ell_j}(t_j - \e s_j)}{\e}=
s_j u_{\ell_j}(t_j - \e s_j)
\to s_j \left( \bar u_{\ell_j}(t_j) + \frac {\sigma_j}2
k_j \right)
\cr&
\frac {y_{j+1}(t_j)- y_{j+1}(t_j - \e s_j)}{\e}=
s_j u_{j+1}(t_j - \e s_j)
 \to s_j \left( \bar u_{j+1}(t_j) - \frac {\sigma_j}2 \right)
k_j \; .
}
$$
%%%%%%%%%%%%%%%%%%%%%%%%%%%%%%%%%%%%%%%%%%%%%%%%%%%%%%%%%%%%
Thus, from \equ(1.22),
we
obtain
%%%%%%%%%%%%%%%%%%%%%%%%%%%%%%%%%%%%%%%%%%%%%%%%%%%%%%%%%%%%
$$
\eqalign{&
\lim_{\e\to 0} \Cal
T^\e(t_1,\dots,t_n;\ell_1,\dots,\ell_n)=\cr&
(-i)^{2n}\sum_{\sigma_1,\dots,\sigma_n,\sigma'_1,\dots,\sigma'_n=\pm
1}\prod_{j=1}^n\sigma_j\sigma'_j
 \int dV\int {dK\over (2\pi)^{3n}}
\int_0^{+\infty} ds_1 \; \dots
\int_0^{+\infty} ds_n\cr&
\int dX\int {d\Xi\over
(2\pi)^{3n}}\prod_{j=1}^n|\hat \phi(k_j)|^2
f_{n+1}(\bar y_1,\dots,\bar u_{n+1};\bar u_1,\dots,\bar y_{n+1})
\cr&
\prod_{j=1}^n
\exp i\big[ {\xi_j}\cdot (\bar y_{\ell_j}(t_j)- x_{j+1})
-{k_j}\cdot[(\bar w_{j}+k_j\sigma_j)\tau'_j]\big]
\; ,
}
\Eq(1.43)
$$
%%%%%%%%%%%%%%%%%%%%%%%%%%%%%%%%%%%%%%%%%%%%%%%%%%%%%%%%%%%%
where $\bar w_j= \bar u_{\ell_j}(t_j) - \bar u_{j+1}(t_j)$.
As indicated at the beginning of the proof, we have now come up
with factors of the form \equ(prot). Hence, as in the previous section,
there only remains to take care of the $d\Xi$ and $ds_1 \; \dots ds_n$ integration
as we did in \equ(protlim).

\medskip

\noi
The
integration on
$\Xi$ produces a delta function concentrating on the set $\{X\,|\,
x_{j+1}= \bar
y_{\ell_j}(t_j), j=1,\dots,n\}$, meaning that the particle $j+1$ is born at
time $t_j$
exactly where the ancestor is at time $t_j$.
Let $\{\tilde y_j(t), j=1,\dots, n+1\}$ be the trajectories $\bar y_j(t)$
satisfying this
additional condition.
Since such trajectories do not depend on
the $s$'s, we
conclude that
%%%%%%%%%%%%%%%%%%%%%%%%%%%%%%%%%%%%%%%%%%%%%%%%%%%%%%%%%%%%
$$
\eqalign{&
\lim_{\e\to 0} \Cal
T^\e(t_1,\dots,t_n;\ell_1,\dots,\ell_n)=
\cr&
(-i)^{2n}\sum_{\sigma_1,\dots,\sigma_n,\sigma'_1,\dots,\sigma'_n=\pm
1}
\prod_{j=1}^n\sigma_j\sigma'_j\int {dK\over (2\pi)^{3n}}\int dV
\prod_{j=1}^n|\hat \phi(k_j)|^2
\cr&
f_{n+1}(\bar y_1,\dots,\bar y_{n+1};
\bar u_1,\,\dots,\,\bar u_{n+1})
\prod_{j=1}^n
\int_0^{+\infty} ds
\exp\big[-i
s {k_j}\cdot[(\bar w_j+k_j\sigma_j)]
\big]
\; ,
}
\Eq(1.44)
$$
%%%%%%%%%%%%%%%%%%%%%%%%%%%%%%%%%%%%%%%%%%%%%%%%%%%%%%%%%%%%
where the integrals on the $s$'s make sense as distributions
(see \equ(protman) and the bound \equ(bdosc)).

\medskip

\noi There remains to explicitly compute the integral over $s$ and
the sum over the $\sigma$'s, so as to identify $\Cal
T(t_1,\dots,t_n;\ell_1,\dots,\ell_n)$ in \equ(1.44). Using formula
\equ(vpde) together with the fact that the expression for the
limit is invariant under the transformation
%%%%%%%%%%%%%%%%%%%%%%%%%%%%%%%%%%%%%%%%%%%%%%%%%%%%%%%%%%%%
$$
s \to -s,\quad k_j\to -k_j,\quad \sigma_j\to -\sigma_j,\quad
\sigma'_j\to
-\sigma'_j,
$$
%%%%%%%%%%%%%%%%%%%%%%%%%%%%%%%%%%%%%%%%%%%%%%%%%%%%%%%%%%%%
we can substitute
%%%%%%%%%%%%%%%%%%%%%%%%%%%%%%%%%%%%%%%%%%%%%%%%%%%%%%%%%%%%
$$
\int_0^{+\infty} ds
\exp\big[-i s {k_j}\cdot[(\bar w_j+k_j\sigma_j)]\big]
 \phantom{aaa} \text{ with its real part, i.e. } \phantom{aaa}
\pi \delta \left( k_j \cdot
(\bar w_j+k_j\sigma_j) \right).
$$
%%%%%%%%%%%%%%%%%%%%%%%%%%%%%%%%%%%%%%%%%%%%%%%%%%%%%%%%%%%%
Doing the change of variables $\eta_j= -\sigma_j k_j$, and
$\sigma_j'=-\sigma_j \bar \sigma_j$, the recursive relations \equ(1.42.10)
become
%%%%%%%%%%%%%%%%%%%%%%%%%%%%%%%%%%%%%%%%%%%%%%%%%%%%%%%%%%%%
$$
\eqalign{&
\bar u_{j+1}(t_{j+1}) = \bar u_{j+1}(t_j) + \frac { 1 + \bar \sigma_j}2
\eta_j,
\cr&
\bar u_{\ell_j}(t_{j+1}) =\bar u_{\ell_j}(t_j) -
 \frac { 1 + \bar \sigma_j}2
\eta_j
\; .
}
\Eq(1.44.10)
$$
%%%%%%%%%%%%%%%%%%%%%%%%%%%%%%%%%%%%%%%%%%%%%%%%%%%%%%%%%%%%
Then the sum on $\sigma_j$ is free and Eq. \equ(1.44) becomes
%%%%%%%%%%%%%%%%%%%%%%%%%%%%%%%%%%%%%%%%%%%%%%%%%%%%%%%%%%%%
$$
\eqalign{&
\lim_{\e\to 0} \Cal
T^\e(t_1,\dots,t_n;\ell_1,\dots,\ell_n)=
%\cr&
\sum_{\bar \sigma_1,\dots,\bar \sigma_n=\pm
1}\prod_{j=1}^n\bar \sigma_j\int {d\eta_1\cdots
d\eta_n \over (2\pi)^{2n}}\int dV
\prod_{j=1}^n|\hat \phi(\eta_j)|^2
\cr&
\qquad
\qquad
f_{n+1}(\tilde y_1,\dots,y_{n+1};
\bar u_1,\,\dots,\,\bar u_{n+1})
\delta\left(
{\eta_j}\cdot(\bar w_j-\eta_j)\right) \; .
}
\Eq(1.44.11)
$$
%%%%%%%%%%%%%%%%%%%%%%%%%%%%%%%%%%%%%%%%%%%%%%%%%%%%%%%%%%%%
We perform partially the integration in $\eta_j$ in the following way:
let $\gamma$ be a smooth function, we have, going to polar coordinates,
%%%%%%%%%%%%%%%%%%%%%%%%%%%%%%%%%%%%%%%%%%%%%%%%%%%%%%%%%%%%
$$
\eqalign{&
\int d\eta \,\gamma(\eta) \,\delta(\eta \cdot (w -\eta)) =
\int_{{\Bbb S}_2}d\o \int_0^{+\infty} d\lambda \,\lambda^2
\gamma( \lambda \o) \delta ( \lambda \o \cdot w - \lambda^2)
\cr&
\qquad
=
\int_{{\Bbb S}_2}d\o\, (\o \cdot w )\,
\gamma
((\o \cdot w) \o)
\text{\bf 1}\{ \o \cdot w > 0 \}
\cr&
\qquad
=
\frac 12 \int_{\Bbb S_2} d\o \, |\o \cdot w| \,\gamma((\o \cdot w) \o) \;
. }
$$
%%%%%%%%%%%%%%%%%%%%%%%%%%%%%%%%%%%%%%%%%%%%%%%%%%%%%%%%%%%%
Inserting this identity in \equ(1.44.10),
we recover the classical trajectories
defined in Eq.s \equ(1.12.1)-\equ(1.12.3). Doing the integrations
in $\lambda_j = |\eta_j|$, we identify the limit in Eq. \equ(1.44.11)
with Eq. \equ(1.12), where
%%%%%%%%%%%%%%%%%%%%%%%%%%%%%%%%%%%%%%%%%%%%%%%%%%%%%%%%%%%%
$$
B(\omega,w) = \frac {1}{8\pi^2} |\o \cdot w| \,\left|
\hat \phi((\o\cdot w)\o) \right|^2.
$$
%%%%%%%%%%%%%%%%%%%%%%%%%%%%%%%%%%%%%%%%%%%%%%%%%%%%%%%%%%%%
\qed

\bigskip
\bigskip

\noi We remark that the cross-section we have found in the
limiting Boltzmann equation is that given by the Born
approximation for the Quantum Scattering problem. This is known as
Fermi's Golden Rule (see [AM], [RV], [Bo], [CTDL], [Co] ...). For
the low-density limit the situation is different. Now all terms in
the perturbative expansion of $S_{int}(t)$ play the same role: We
thus expect that the same result holds as Theorem 3.1, but with a
cross section given by the full scattering matrix $\Cal S$
associated to the potential $\phi$. We mention in passing the
following well-known fact [RS]: $\Cal S$ admits a power series
expansion in the potential $\phi$, called the Born series
expansion, whose first term is precisely given by the Fermi Golden
Rule,
%%%%%%%%%%%%%%%%%%%%%%%%%%%%%%%%%%%%%%%%%%%%%%%%%%%%%%%%%%%%
$$
\Cal S(k)
=
|\hat\phi(k)|^2+O\(\hat\phi^3\) \; ,
$$
%%%%%%%%%%%%%%%%%%%%%%%%%%%%%%%%%%%%%%%%%%%%%%%%%%%%%%%%%%%%
roughly. The weak coupling regime may thus be seen (technically) as
some first order approximation of the low density regime.
We refer to [Ni1], [Ni2] for a kinetic
interpretation of the scattering matrix in a semi-classical regime.
We also refer to [Ca1], [Ca2], for the derivation of a linear
Boltzmann equation involving the Born series expansion
in a low density regime. We finally quote [Co] for a physical
discussion of related questions.

\medskip

\noi

Some comments are in order.

We underline once more that while in the low-density regime
classical and quantum systems evolve similarly according to the
Boltzmann equation, the situation changes drastically in the
weak-coupling limit. Here contrary to the behavior outlined in the
present paper for quantum systems, classical systems of particles
are expected to satisfy the Landau equation which a diffusive
character (see [Sp2]). Unfortunately no rigorous result is known
in this direction. The situation is better understood for linear
problems. Obviously the same scalings can be also considered for a
test particle moving in a random distribution of obstacles. In
this context it is possible to derive a linear transport equation
for a classical particle in the low-density regime (see [G],
[BBS], the review paper [Sp3], or the textbook [Sp2]) or the
linear Landau equation in the weak coupling limit (see [DGL]). A
linear Boltzmann equation can be derived in the weak-coupling
limit either for short times (see Refs [Sp1],[La], [HLW] ) or
globally in time (see [EY1], [EY2]). The one-dimensional case is
somehow pathological (see [EPT]). For the case of non random
scatterers negative results are available [CP1], [CP2] (see also
[BGW] in the classical context).

Similar considerations in the case of an atom coupled to a gas have been
developed in [D\"u]. In addition, notice that the problem of the wave
motion in a random medium  is of interest for the applications as shown
in Ref.s [KPR] We finally mention [PV] for the analysis of a weak
coupling regime when the obstacles are temporally random (and the
underlying process is at once (almost) Markovian).

At the physical level, the question of
passing from the Schr\"odinger equations to (linear or non-linear)
Boltzmann equations is an old problem.
We may quote
[Pa], [KL1], [KL2],
[Ku], and [VH1], [VH2], [Zw], as well as the textbooks quoted before.

When this paper was finished we were informed by H.T. Yau that
arguments similar to those of the present paper have been
independently developed by him and his group in a paper in
preparation.

\bigskip

\centerline{\bf REFERENCES}

\bigskip

\noi
[AM] N.W. Ashcroft, N.D. Mermin,
{\bf Solid stats physics}, Saunders, Philadelphia (1976).

\noi
[Bo] A. Bohm, {\bf
Quantum Mechanics}, Texts and mo\-no\-graphs in Phy\-sics,
Sprin\-ger-Verlag (1979).

\noi
[BBS] C. Boldrighini, L.A. Bunimovich, Ya. Sinai,
{\it
On the Boltzmann equation for the Lorentz gas},
J. Stat. Phys., Vol. 32, N. 3, pp. 477-501 (1983).

\noi
[BGW] J. Bourgain, F. Golse,
B. Wennberg, {\it On the distribution of free path lengths for
the periodic Lorentz gas},
Comm. Math. Phys., Vol. 190, pp. 491-508 (1998).

\noi
[Ca1]
F. Castella, {\it From the von Neumann equation
to the Quantum Boltzmann equation in a deterministic framework},
J. Stat. Phys., Vol. 104, N. 1/2, pp. 387-447 (2001).

\noi
[Ca2]
F. Castella,
{\it
>From the von Neumann equation
to the Quantum Boltzmann equation II:
identifying the Born series},
J. Stat. Phys., Vol. 106, N. 5/6, pp. 1197-1220 (2002).

%\noi
%[Ca3]
%F. Castella,
%{\it
%R\'esultats de convergence et de non-convergence de l'\'equation
%de von Neumann p\'eriodique vers l'\'equation de Boltzmann quantique},
%S\'eminaire E.D.P., Ecole Polytechnique,
%expos\'e N. XXI, 1999-2000.

\noi
[CP1] F. Castella, A. Plagne,
{\it A distribution result for slices
of sums of squares},
Math. Proc. Cambridge Philos. Soc.,
Vol. 132, N.1, pp. 1-22 (2002).

\noi
[CP2] F. Castella, A. Plagne,
{\it Non-derivation
of the Quantum Boltzmann equation from the periodic Schr\"odinger equation},
To appear in Indiana Univ. Math. J. (2003).

\noi
[CC]
S. Chapman and T. G. Cowling, {\bf The Mathematical
Theory of
Non-uniform
Gases}, Cambridge Univ. Press, Cambridge, England (1970).

\noi
[CIP]
C. Cercignani, R. Illner, M. Pulvirenti,
{\bf The mathematical theory of dilute gases},
Applied Mathematical Sciences, Vol. 106,
Springer-Verlag, New York (1994).

\noi
[Ch] S.L. Chuang, {\bf
Physics of optoelectronic
devices}, Wiley series in pure and applied optics,
New-York (1995).

\noi
[CTDL] C. Cohen-Tannoudji, B. Diu, F. Lalo\"e,
{\bf M\'ecanique Quantique},
I et II, Enseignement des Sciences, Vol. 16, Hermann (1973).

%\noi
%[CTDRG]
%C. Cohen-Tannoudji, J. Dupont-Roc, and G. Grynberg,
%{\bf Processus d'inter\-ac\-ti\-on entre photons et atomes},
%"Savoirs actuels", In\-ter\-e\-di\-ti\-ons/E\-di\-ti\-ons du CNRS (1988).

\noi
[Co] M. Combescot,
{\it
On the generalized golden rule for transition probabilities},
Phys. A: Math. Gene., Vol. 34, N. 31, pp. 6087-6104 (2001).

\noi
[D\"u]
R. D\"umcke,
{\it The low density limit for an $N$-level system
interacting with a free Bose or Fermi gas},
Comm. Math. Phys., Vol. 97, N. 3, pp. 331-359 (1985).

\noi
[DGL]
D. D\"urr, S. Goldstein, J.L. Lebowitz
{\it Asymptotic motion of a classical particle in a random potential
in two dimension: the Landau model},
Comm. Math. Phys., N. 113,
pp. 209-230 (1987).

\noi
[EY1] L. Erd\"os, H.T. Yau, {\it
Linear Boltzmann Equation as Scaling Limit of Quantum Lorentz Gas}
Advances in differential equations and mathematical physics (Atlanta,
GA, 1997), pp. 137-155, Contemp. Math., 217, Amer. Math. Soc., Providence, RI (1998).

\noi
[EY2] L. Erd\"os, H.T. Yau, {\it
Linear Boltzmann equation as the weak coupling limit of
a random Schr\"odinger equation},
Comm. Pure Appl. Math., Vol. 53, N. 6, pp. 667-735
(2000).

\noi
[EPT]
R. Esposito, M. Pulvirenti, A. Teta,
{\it
The Boltzmann equation for a one-dimensional quantum Lorentz
gas},
Comm. Math. Phys., Vol.  204, no. 3, pp. 619-649 (1999),
and
{\it Erratum: "The Boltzmann equation for a one-dimensional quantum
Lorentz gas"},
Comm. Math. Phys., Vol.  214, N.
2, pp. 493-494 (2000).

\noi
[Fi] M.V. Fischetti, {\it Theory of electron transport
in small semiconductor devices using the Pauli master equations},
J. Appl. Phys., Vol. 83, N. 1, pp. 270-291 (1998).

\noi
[G]
G. Gallavotti,
{\it Time evolution problems in classical statistical mechanics and the
wind-tree-model} Cargese Lectures in Physics, vol IV, ed. D. Kastler,
Gordon Breach, Paris, (1970);{\it Divergences and approach to equilibrium
in the Lorenz and the  wind-tree models} Phys.Rev. {\bf 185}, 308--322
(1969). See also the book {\bf Statistical Mechanics},
Appendix A2 to Ch. 1,  Springer--Verlag (1999).

\noi
[HLW] T.G. Ho, L.J. Landau, A.J. Wilkins,
{\it On the weak coupling limit for a Fermi gas in a random potential},
Rev. Math. Phys., Vol. 5, N. 2, pp. 209-298 (1993).

\noi
[H\"o] L. H\"ormander,
{\bf The analysis of linear partial differential operators},
Springer-Verlag, Berlin (1994).

\noi
[Hu]
K. Huang, {\bf
Statistical mechanics}, Wiley and Sons (1963).

\noi
[IP]
R. Illner, M. Pulvirenti, {\it Global Validity of the Boltzmann
equation
for a two dimensional rare gas in vacuum},
Comm. Math. Phys., Vol. 105, pp. 189-203 (1986).
{\it Erratum and improved result}
Comm. Math. Phys., Vol. 121, pp. 143-146 (1989).

\noi
[KPR] J.B. Keller, G. Papanicolaou,
L. Ryzhik, {\it Transport equations for elastic and other waves
in random media},  Wave Motion, Vol. 24,
N. 4, p. 327-370 (1996).

\noi
[KL1]
W. Kohn, J.M. Luttinger, Phys. Rev., Vol. 108, pp. 590 (1957).

\noi
[KL2] W. Kohn, J.M. Luttinger,
Phys. Rev., Vol. 109, pp. 1892 (1958).

\noi
[Ku] R. Kubo, J. Phys. Soc. Jap., Vol. 12 (1958).

\noi
[L]
O. Lanford III,
{\bf Time evolution of large classical systems},
Lecture Notes in Physics, Vol. 38, pp. 1-111, E.J. Moser ed.,
Springer-Verlag (1975).

\noi
[La] L.J. Landau, {\it Observation of Quantum Particles on a
Large Space-Time Scale}, J. Stat. Phys., Vol. 77, N. 1-2, pp. 259-309 (1994).

\noi
[MRS]
P.A. Markowich, C. Ringhofer, C. Schmeiser,
{\bf Semiconductor equations},
Sprin\-ger-Verlag, Vienna  (1990).

\noi
[Ni1] F. Nier, {\it Asymptotic Analysis of a scaled Wigner equation and
Quantum Scattering}, Transp. Theor. Stat. Phys.,
Vol. 24, N. 4 et 5, pp. 591-629 (1995).

\noi
[Ni2] F. Nier, {\it A semi-classical picture of quantum scattering},
Ann. Sci. Ec. Norm. Sup., 4. S\'er., t. 29, p. 149-183 (1996).

\noi
[PV]
F. Poupaud, A. Vasseur,
{\it
Classical and quantum transport in random media},
Preprint University of Nice (2001).

\noi
[Pa] W. Pauli, {\it Festschrift
zum 60. Geburtstage A. Sommerfelds}, p. 30, Hirzel, Leipzig (1928).

\noi
[RS]
M. Reed, B. Simon,
{\bf Methods of modern mathematical physics III. Scattering theory},
Academic Press,
New York-London ( 1979).

\noi
[RV]
E. Rosencher, B. Vinter, {\bf Optoelectronique},
Dunod (2002).

\noi
[Sp1] H. Spohn, {\it Derivation of the transport equation
for electrons moving through random impurities}, J. Stat. Phys.,
Vol. 17, N. 6, pp. 385-412 (1977).

\noi
[Sp2] H. Spohn, {\bf Large scale dynamics of interacting particles},
Springer (1991).

\noi
[Sp3] H. Spohn,
{\it Kinetic equations from Hamiltonian dynamics: Markovian limits},
Rev. Modern Phys., Vol.  52, N. 3, pp. 569-615 (1980).

\noi
[VH1] L. Van Hove, Physica, Vol. 21 p. 517 (1955).

\noi
[VH2] L. Van Hove, Physica, Vol. 23 p. 441 (1957).

\noi
[VH3] L. Van Hove, in {\bf Fundamental Problems
in Statistical Mechanics}, E.G.D. Cohen ed., p. 157 (1962).

\noi
[Zw] R. Zwanzig, {\bf Quantum Statistical Mechanics},
P.H.E. Meijer ed., Gordon and Breach, New-York (1966).

\end